\documentclass[journal,12pt,onecolumn,draftclsnofoot,]{IEEEtran}
\usepackage{multicol}   
\usepackage{multirow}
\usepackage[pdftex]{graphicx}
\usepackage[ruled]{algorithm2e}
\usepackage{algorithmic}
\usepackage[small]{caption}
\usepackage{float}
\usepackage{amssymb,amsmath}
\usepackage{graphicx}
\usepackage{subfigure}
\usepackage{xspace}
\usepackage{amsthm}
\usepackage{fancyhdr}

\usepackage{caption,setspace}
\usepackage{amssymb,amsmath}

\usepackage[top=1 in, bottom=0.7 in, left=0.6 in, right=0.6 in]{geometry}
\usepackage{subfig}
\usepackage{textcomp} 
\usepackage{epstopdf}
\usepackage{bbding}
\usepackage{cite} 
\usepackage{color}
\usepackage{pifont}
\usepackage{lipsum}

\begin{document}

\title{\huge Energy-Efficient Backscatter-Assisted Coded Cooperative-NOMA for B5G Wireless Communications }
\author{Muhammad Asif, Asim Ihsan, Wali Ullah Khan,  \IEEEmembership{Member, IEEE},  Ali Ranjha, Shengli Zhang, \IEEEmembership{Senior Member, IEEE}, and Sissi Xiaoxiao Wu, \IEEEmembership{Member, IEEE}, \thanks{Muhammad Asif, Shengli Zhang, and Sissi Xiaoxiao Wu are with the Guangdong Key Laboratory of Intelligent Information Processing, College of Electronics and Information Engineering, Shenzhen University, Shenzhen, Guangdong 518060, China. (emails: masif@szu.edu.cn, zsl@szu.edu.cn, xxwu.eesissi@gmail.com).

Asim Ihsan is with the School of Computer Science and Electronic Engineering, Bangor University, Bangor LL57 1UT, U.K. (e-mail:
a.ihsan@bangor.ac.uk)
		
Wali Ullah Khan is with the Interdisciplinary Centre for Security, Reliability and Trust (SnT), University of Luxembourg, 1855 Luxembourg City, Luxembourg (Emails: waliullah.khan@uni.lu, waliullahkhan30@gmail.com).

Ali Ranjha is with the Department of Electrical Engineering, École de Technologie Supérieure, Montréal, Quebec, Canada, (email: ali- nawaz.ranjha.1@ens.estmtl.ca).

(Corresponding author: Sissi Xiaoxiao Wu.)
}

\vspace{-0.6cm}}%

\markboth{Submitted to IEEE}%
{Shell \MakeLowercase{\textit{et al.}}: Bare Demo of IEEEtran.cls for IEEE Journals} 

\maketitle

\begin{abstract} 
In this manuscript, we propose an alternating optimization framework to maximize the energy efficiency of a backscatter-enabled cooperative Non-orthogonal multiple access (NOMA) system by optimizing the transmit power of the source, power allocation coefficients (PAC), and power of the relay node under imperfect successive interference cancellation (SIC) decoding. A three-stage low-complexity energy-efficient alternating optimization algorithm is introduced which optimizes the transmit power, PAC, and relay power by considering the quality of service (QoS), power budget, and cooperation constraints. Subsequently, a joint channel coding framework is introduced to enhance the performance of far user which has no direct communication link with the base station (BS) and has bad channel conditions. In the destination node, the far user data is jointly decoded using a Sum-product algorithm (SPA) based joint iterative decoder realized by jointly-designed Quasi-cyclic Low-density parity-check (QC-LDPC) codes. Simulation results evince that the proposed backscatter-enabled cooperative NOMA system outperforms its counterpart by providing an efficient performance in terms of energy efficiency. Also, proposed jointly-designed QC-LDPC codes provide an excellent bit-error-rate (BER) performance by jointly decoding the far user data for considered BSC cooperative NOMA system with only a few decoding iterations.   
\end{abstract}


\IEEEpeerreviewmaketitle

\section{Introduction}
The upcoming beyond fifth-generation (B5G) communication networks have become the focal choice for researchers across academia and industry because of their capability to support billions of communication devices \cite{giordani2020toward}. Furthermore, internet-of-things (IoT) has been considered as a new technological revolution by providing services in a broad spectrum of applications including smart homes, smart cities, autonomous vehicles, smart ports, smart hospitals, unmanned aerial vehicles (UAV), and so on \cite{jameel2020noma,liu2019next}. However, this anomalous revolution in next-generation communication systems would require efficient utilization of available resources in terms of spectrum and energy \cite{khan2019joint,tanveer2021enhanced}. Furthermore, due to limited energy resources for IoT sensors deployed in mountains, dense forests, radioactive fields, and hidden in appliances and walls of smart buildings, it would be very difficult and cost-effective to replace their batteries after a regular interval of time \cite{jameel2019applications,khan2021noma}. In such kinds of scenarios, it is highly desired to find some flexible techniques to enhance the life period of sensors without replacing their batteries \cite{lu2018ambient,khan2021energy}. Due to the low energy consumption property of IoT sensor devices, ambient energy harvesting seems to be a promising approach to enhance the life of a sensor node. In this regard, ambient backscatter communication (BSC) has attracted the attention of researchers because of its capability to operate at a very low-power level and harvests energy from existing radio signals available in the surrounding environment \cite{ihsan2021energy}. The ambient backscatter modulates and reflects the incoming signals towards different users by utilizing the harvested power from existing radio signals \cite{van2018ambient}. Furthermore, ambient backscatters could be very efficient to enhance the coverage and capacity of the network when there is no direct path between base station (BS) and cell-edge distant users \cite{khan2021learning}.  

To fulfill the spectrum and energy requirements for next-generation communication systems, Non-orthogonal multiple access (NOMA) and backscatter communication are two emerging technologies for a wide range of applications. It has also been shown that NOMA outperforms its orthogonal-multiple access (OMA) counterpart in terms of sum capacity  (sum-rate), spectral and energy efficiency, and massive connectivity \cite{khan2019joint,li2021physical}. In the NOMA protocol, multiple users are served simultaneously over the same frequency/code where multiplexing is achieved by providing different power levels to different users. To achieve user fairness, less power is allocated to near users with good channel gains, and more power is allocated to far users with bad channel conditions. To enable NOMA transmission for different users, two technologies called superposition coding (SC) and successive-cancellation decoding (SIC) are deployed at the transmitter and receiver, respectively \cite{makki2020survey,ding2017application}.        
  
\subsection{Technical Literature Review}
Various studies have been conducted by integrating the backscatter communication with the OMA protocol. For instance, the achievable rates and closed-form expressions for optimal power allocation using cooperative backscatter communication network are provided by Guo {\em et al.} \cite{8692391}. Yinghui {\em et al.} investigated the performance of ambient backscatter in terms of outage probability (OP) and proposed an adaptive reflection coefficient which significantly reduces the outage probability \cite{9051982}. The authors of \cite{jameel2019simultaneous} have computed the closed-form expressions for ambient backscatter which offers a better trade-off between harvested energy and achievable rate under the Rayleigh-fading channel. Wang {\em et al.} proposed a resource allocation framework for backscatter communication networks to minimize the energy consumption of mobile users \cite{wang2021joint}. Authors in \cite{zhu2021distributed}, proposed a distributed resource allocation framework for channel selection and optimal power allocation for ambient backscatter. An optimal energy detector and expression for symbol-error rate (SER) of ambient backscatter communication network have been provided in \cite{8423609}. Authors of \cite{8093703} proposed an optimization framework for throughput maximization which provides an optimal trade-off between the active and sleep states and reflection coefficient of ambient backscatter.

Recently, several efforts have been utilized to combine backscatter communication with NOMA to enhance the throughput, spectral efficiency, coverage, and connectivity of the system. For example, Khan {\em et al.} \cite{khan2021backscatter123} investigated a joint optimization problem for NOMA backscatter-enabled vehicle-to-everything (V2X) communication network by efficiently optimizing the power allocation of base BS and roadside units. Authors of \cite{khan2021backscatter} have computed the closed-form expressions for optimal power allocation of BS and reflection coefficient of backscatter to maximize the sum-rate of backscatter-enabled NOMA network. In \cite{zhang2019backscatter}, the performance of the NOMA-enabled BSC ambient cellular network has been investigated by providing closed-form solutions for the OP and ergodic capacity. Nazar {\em et al.} \cite{nazar2021ber} have computed the closed-form expressions in terms of bit-error-rate (BER) for NOMA-enabled BSC system. An energy efficiency (EE) maximization problem for the backscatter NOMA system has been solved by optimizing the BS transmit power and reflection coefficient of backscatter in \cite{xu2020energy}. Authors of \cite{li2021hardware} have investigated the physical-layer security of back-scatter NOMA in terms of security and reliability. They have also computed the analytical solutions for the OP and intercept probability. Further, a backscatter NOMA-enabled hybrid technique is proposed in \cite{zeb2019noma} to improve system performance by decreasing the OP and increasing the system throughput. Yang{\em et al.} \cite{yang2019resource} investigated an optimization problem for backscatter NOMA to maximize the system throughput by optimizing the time and reflection coefficient of the backscatter device. To maximize the throughput, the work of \cite{liao2020resource} proposed a resource allocation framework for a NOMA-enabled full-duplex symbiotic radio system.  Besides, Li {\em et al.} \cite{li2020secrecy} discussed the security issues of ambient backscatter NOMA system and provided the analytical expressions of the OP and intercept probability. Further, Authors of \cite{farajzadeh2019uav} have presented a resource allocation framework to minimize the flight time and maximize the number of decoded bits.

Furthermore, cooperative communication has the potential to improve the performance and coverage of next-generation communication systems. Authors in \cite{ali2021deep} provided a power optimization framework to maximize the sum-rate of cooperative NOMA. To achieve fairness among secondary users in cooperative NOMA-enabled cognitive radio transmission, the power allocation optimization problem has been investigated in \cite{ali2022fair}. The works of \cite{jiang2018adaptive} have provided an optimization setup for a cooperative NOMA-assisted device-to-device communication network to achieve the maximum sum-rate of the system. Also, authors of \cite{bae2019joint} derived closed-form solutions for optimal power and time allocation for cooperative NOMA system. Power allocation and power splitting coefficients have been optimized for cooperative NOMA-enabled energy harvesting system in \cite{van2018analysis}. The authors of \cite{jameel2019secrecy} have discussed the security issues of cooperative NOMA energy harvesting system. In \cite{ning2019energy}, a stochastic energy-efficiency problem has been investigated to provide a better trade-off between the delay and energy efficiency for a cooperative NOMA network. Further, for cooperative NOMA, closed-form expressions for ergodic sum-capacity and outage probability have been presented in \cite{kara2020threshold}. In addition, authors of \cite{chen2021backscatter} have investigated a power allocation problem to maximize the expected rate of backscatter-enabled cooperative NOMA system.

\subsection{ Research Motivation and Contributions}
Most of the existing literature \cite{xu2020energy}, \cite{yang2019resource}, \cite{liao2020resource}  \cite{ali2021deep,ali2022fair,jiang2018adaptive,bae2019joint,van2018analysis,jameel2019secrecy,ning2019energy,kara2020threshold,chen2021backscatter} presented above has considered the ideal situation by assuming perfect SIC decoding for different NOMA-enabled communication systems which is not practical in real-life scenario. Furthermore, it is very difficult to completely remove the interference caused by other users while considering the imperfect SIC decoding at the receiver. Consequently, this residual proportion of signals power from interfering users significantly degrades the performance of the system in terms of bit-error rate, sum-rate, and energy efficiency. Moreover, in cooperative NOMA, the relay node exploits the SIC decoding principle to estimate the far user's data. However, if the relay node fails to decode the data for the far user and a decoding error occurs, then this error will propagate and the performance of both relay and destination node would be compromised. Hence, to avoid this chain process of error propagation, efficient error-correction codes can be utilized at the relay and destination nodes \cite{asif2020jointly, tang2016joint, zhang2016joint}.

However, some of the researchers \cite{khan2021noma,khan2021backscatter, nazar2021ber,li2021hardware} have assumed the imperfect SIC decoding for NOMA communication systems, but, the cooperation among near and far users has not been considered. Further, in a real-life scenario, it could be very common when there is no direct path between the BS and far user of the NOMA network. In this situation, backscatter can play a crucial role to reflect the modulated signal towards the near and far user in the first time slot of cooperative NOMA which also increases the coverage of BS. In the second time frame, the near user can work as a  relay node to broadcast the data towards the far user, thereby increasing the diversity and capacity of the system. Moreover, the works of \cite{khan2021noma,khan2021backscatter} have considered the imperfect SIC decoding at the receiver, however, only transmit power is optimized to maximize the sum-rate, but, cooperation among NOMA users has not been considered.

Based on the above discussion, it is highly desired to investigate the backscatter-enabled cooperative NOMA system under imperfect SIC decoding at the receiver assuming no direct link between BS and far user. To the best of our knowledge, an optimization framework that simultaneously optimizes the transmit power of the source, power allocation coefficients (PAC) of NOMA users, and power of relay node to maximize the energy efficiency (EE) by considering SIC decoding under the uncertainties for backscatter-enabled cooperative NOMA system has not been investigated yet. To address this gap, this work aims to propose a new optimization framework that simultaneously optimizes the transmit power of source, PAC of NOMA users, and power of relay node to maximize the EE by considering the QoS, power budgets, and cooperation constraints under the imperfect SIC decoding for BSC-enabled cooperative NOMA system. The proposed optimization problem has been transformed from fractional to substantive form using Dinkelback method \cite{dinkelbach1967nonlinear}, then, the transformed problem is decoupled into three sub-problems to find the closed-form expressions. Finally, the closed-form solutions in terms of transmit power for source node, PAC, and power of relay nodes have been computed based on Karush-Kuhn-Tucker (KKT) and sub-gradient methods. The main contributions of this manuscript are summarized as follows:        

\begin{enumerate} 
\item In PD-NOMA communications, the performance of distant cell-edge users is largely affected due to bad channel conditions, which ultimately degrades the overall energy efficiency of the system. Thus, an ambient-backscatter enabled cooperative NOMA network has been considered to improve the performance of the system, in terms of energy-efficiency, where a backscatter node has been exploited to improve coverage and capacity of the system.

\item Since, the far user suffers from severe performance degradation due to bad channel conditions. Also, for the considered system, we assume that there is no direct link between BS and far user due to large obstacles. Hence, an ambient backscatter node is exploited which enhances the communication quality and coverage by transmitting the signal towards the far user of the considered system. Moreover, the performance of far user is further improved by incorporating the cooperation among NOMA users.

\item For practical realization of NOMA, the near user applies SIC principle to remove the far user signal before decoding its signal. However, in a real-life scenario, it is very difficult for the near user to completely remove the far user signal. Hence, the imperfect SIC impacts the outage, throughput, and energy-efficiency performance of the system. In this regard, we have analyzed the performance of the considered backscatter-enabled cooperative NOMA system under imperfect SIC decoding at the near user. 

\item Further, to maximize the energy-efficiency of the considered system, a low-complexity energy-efficient alternating optimization algorithm is introduced which optimizes the transmit power of the source, PAC, and power of relay node by considering the QoS, power budget, and cooperation constraints under the imperfect SIC decoding at NOMA receivers.     

\item In cooperative NOMA networks, the relay node exploits the SIC principle to decode the far user's data. However, if the relay node fails to decode the information and a decoding error occurs, then this error will propagate and the performance of both the relay and destination node would be affected. Thus, to further improve the performance of the far user, a channel coding framework has been proposed to jointly decode the far user's data received from the backscatter-aided source and relay node based on the Sum-product algorithm (SPA) aided iterative decoder realized by jointly-designed QC-LDPC codes. Simulation results evince that the proposed jointly-designed QC-LDPC codes provide an efficient error-correction performance by jointly decoding the far user data with only a few decoding iterations.

\end{enumerate}
The remaining part of this manuscript is arranged as follows: In Section II, the proposed system model of BSC cooperative NOMA network and problem formulation for EE maximization are provided. The optimal solutions of formulated EE maximization problem are given in Section III. In Section IV, jointly designed QC-LDPC codes based on CBSEC for BSC cooperative NOMA system are presented. Numerical simulation results and discussions are given in Section V, and finally, the conclusion of this manuscript is presented in Section VI. 
\begin{figure*}[!t]
\centering
\includegraphics [width=0.75\textwidth]{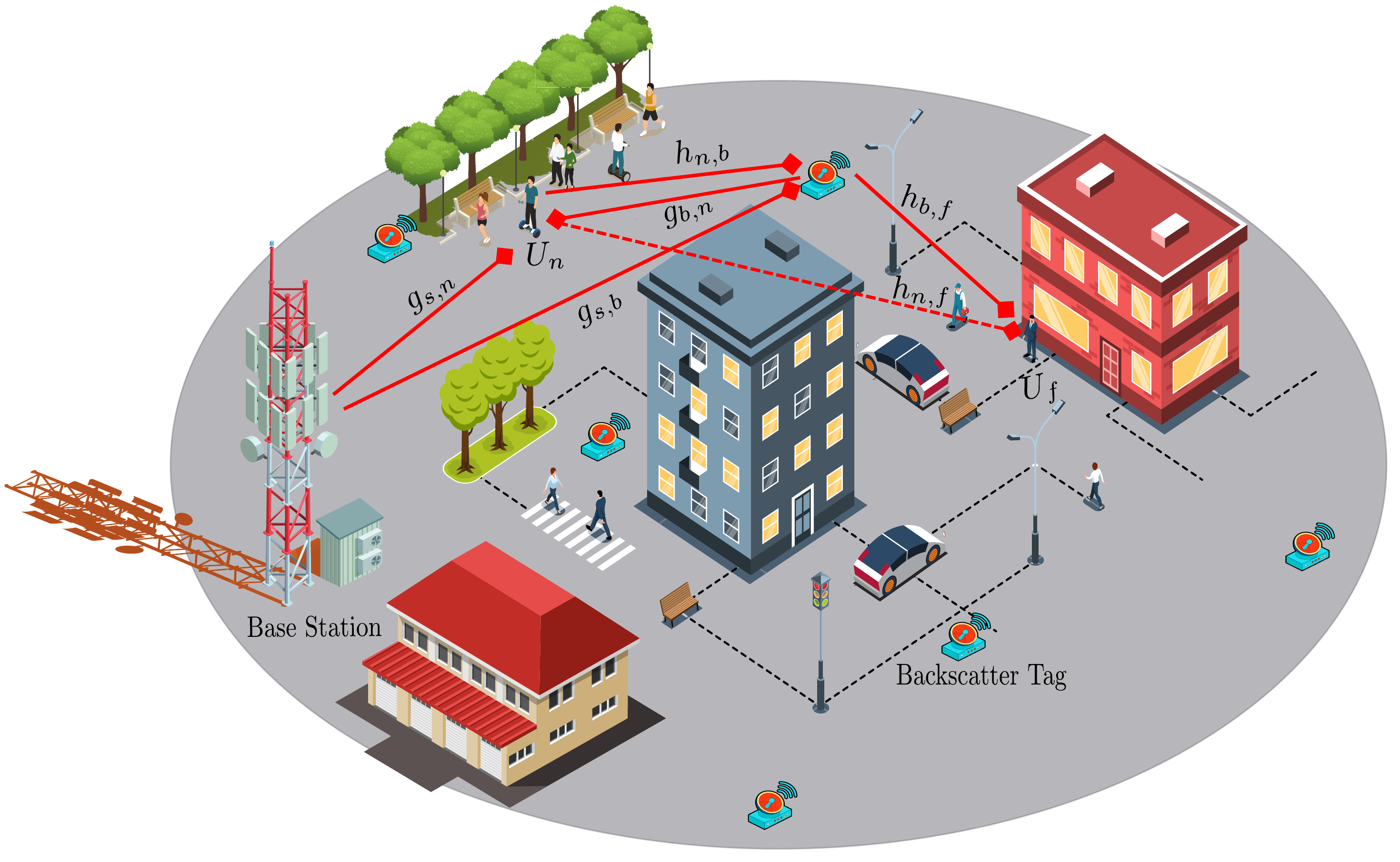}
\caption{Illustration of BSC-enabled cooperative NOMA system model.}
\label{blocky}
\end{figure*}
\section{System Model and Problem Formulation}
A backscatter-enabled cooperative NOMA network considered in this work is depicted in Fig. \ref{blocky}, where a base station is supposed to serve two NOMA users assuming Rayleigh-fading transmission. Moreover, the considered cooperative NOMA system also contains a backscatter tag that modulates and reflects the superimposed data coming from the source node towards both near and far users being served in a downlink scenario. Further, it is assumed that all of the nodes (i.e., BS, near user, and far user) involved in considered BSC cooperative NOMA system are equipped with a single antenna terminal. We assume that the perfect channel state information (CSI) is available at the source node. Based on the NOMA protocol, the user closer to the BS, denoted as $U_{n}$, has better channel condition, whereas the far user with bad channel condition is denoted as $U_{f}$. We also assume that there is no direct communication link between BS and $U_{f}$, where a backscatter tag (BST) is utilized to reflect signal towards $U_{f}$. Further, the SIC process is performed at $U_{n}$ to remove the $U_{f}$ data, but it's very difficult to completely remove the $U_{f}$ data from the received superimposed signal which can cause a SIC error. Therefore, we assumed imperfect SIC decoding at the receiver for the proposed BSC cooperative NOMA network. In the first time frame, BS broadcasts the superimposed data towards $U_{n}$, $U_{f}$, and BST. The BST modulates and reflects the superimposed data coming from BS towards $U_{n}$ and $U_{f}$. Since, the $U_{n}$ receives two copies of superimposed data from BS and BST, respectively. On the other hand, it is important to note that the $U_{f}$ receives the data only from BST because there is no direct link between BS and $U_{f}$. Based on the SIC principle \cite{khan2021noma,khan2021backscatter}, the $U_{n}$ first decodes the $U_{f}$ data and then subtracts this decoded data from the received signal to decode its own information. Consequently, the superimposed signals received at $U_{n}$ and $U_{f}$ during the first time slot are given as follows:      
\begin{align}
y_{n,1}&=\sqrt{g_{s,n}}(\sqrt{P\xi_{n}}x_{n}+ \sqrt{P\xi_{f}}x_{f}) +\nonumber\\ & \sqrt{g_{s,b}g_{b,n}\psi_{i}}(\sqrt{P\xi_{n}}x_{n}+ \sqrt{P\xi_{f}}x_{f})m_b + w_{n,1}, \label{1}
\end{align}
\begin{align}
	y_{f,1}= \sqrt{g_{s,b}h_{b,f}\psi_{i}}(\sqrt{P\xi_{n}}x_{n}+ \sqrt{P\xi_{f}}x_{f})m_b+ w_{f,1}, \label{2}
\end{align}
where $P$ represents the transmit power of BS, $\xi_{n}$ and $\xi_{f}$ denote the power allocation coefficients for near user and far user, respectively. $x_{n}$ and $x_{f}$ are unit variance data symbols for $U_{n}$ and $U_{f}$ transmitted from the BS, and $m_b$ is the signal added by backscatter tag such that $E[|m_b|^2]=1$, where $\psi_{i}$ denotes the reflection coefficient of BST in the first time frame. $w_{n,1}$ and $w_{f,1}$ denote the additive-white Gaussian noises (AWGN) with zero mean and variance $\sigma^{2}$ at $U_{n}$ and $U_{f}$, respectively, during the first time slot. Further, $g_{s,n}$, $g_{s,b}$, and $g_{b,n}$ represent the channel gains from BS to $U_{n}$, BS to BST, and BST to $U_{n}$, respectively. Similarly, $h_{b,f}$ is the channel gain from BST to $U_{f}$ in first time slot of proposed cooperative NOMA system.

In the second time slot, the $U_{n}$ acts as a relay and broadcasts the $U_{f}$ data. The $U_{f}$ data is also received at the BST node which modulates and reflects the received signal towards $U_{f}$. Thus, $U_{f}$ receives two copies of data from $U_{n}$ and BST node, thereby increasing the diversity and capacity of the network. Consequently, the signal received at $U_{f}$ in the second time frame is given as:   
    
\begin{align}
	y_{f,2}= \sqrt{P_{r} h_{n,f}}x_{f}+ \sqrt {P_{r} h_{n,b} h_{b,f}\psi_{j}}x_{f}m_b +w_{f,2}, \label{3}
\end{align}
where $P_{r}$ denotes the transmit power of relay node, $w_{f,2}$ is the AWGN noise at $U_{f}$ with zero mean and variance $\sigma^{2}$ , and $\psi_{j}$ represents the reflection coefficient of BST in the second time frame. Further, $h_{n,b}$, $h_{b,f}$ and $h_{n,f}$ are the channel gains from $U_{n}$ to BST, BST to $U_{f}$, and $U_{n}$ to $U_{f}$, respectively. 

Based on NOMA protocol, the near user $U_{n}$ applies the SIC principle to decode its data $x_{n}$ and backscatter information $m_b$ from the received superimposed signal. On the other hand, far user $U_{f}$ directly decodes its data $x_{f}$ from the superimposed signal by treating the $U_{n}$ and $m_b$ as interfering signals. However, in a practical scenario, it is very difficult for $U_{n}$ to make SIC error equal to zero by completely removing the interference of $U_{f}$ and $m_b$ signals. Consequently, SIC error is probable at $U_{n}$ such that the far user data $x_{f}$ can not be removed completely from the superimposed signal received at the $U_{n}$ in the first time slot. Therefore, the desired signal to interference plus noise ratio (SINR) at $U_{n}$ to decode the $U_{f}$ signal is given as follows:
\begin{align}
\gamma_{nf}=\frac{P\xi_{f}(g_{s,n}+g_{s,b}g_{b,n}\psi_{i})}{P\xi_{n}(g_{s,n}+g_{s,b}g_{b,n}\psi_{i})+\sigma^{2}}, \label{4}
\end{align}
where the corresponding data rate at $U_{n}$ to decode the $U_{f}$ data in first time frame can be expressed as $R_{nf} = \frac{1}{2}W\log_{2}(1+\gamma_{nf})$, $W$ denotes the bandwidth allocated. Further, the SINR at $U_{n}$ to decode its data $x_{n}$ in the first time frame can be expressed as 
\begin{align}
	\gamma_{1}=\frac{P\xi_{n}(g_{s,n}+g_{s,b}g_{b,n}\psi_{i})}{P\xi_{f}(g_{s,n}+g_{s,b}g_{b,n}\psi_{i})\eta+\sigma^{2}}, \label{5}
\end{align}
where $\eta$ denotes the imperfect SIC parameter which occurs due to the fraction of residual component of $U_{f}$ signal \cite{khan2021backscatter}. Thus, the corresponding data rate of $U_{n}$ to decode its own signal can be expressed as $R_{1} = \frac{1}{2}W\log_{2}(1+\gamma_{1})$. Further, the received SINR at $U_{f}$ to decode its signal during the first time frame can be given as
\begin{align}
	\gamma_{2}=\frac{P\xi_{f}(g_{s,b}h_{b,f}\psi_{i})}{P\xi_{n}(g_{s,b}h_{b,f}\psi_{i})+\sigma^{2}}, \label{6}
\end{align}
where the corresponding data rate at $U_{f}$ to decode its signal is expressed as $R_{2} = \frac{1}{2}W\log_{2}(1+\gamma_{2})$. Similarly, the SINR at $U_{f}$ to decode its signal in the second time slot of the BSC-enabled cooperative NOMA system can be expressed as
\begin{align}
	\gamma_{3}=\frac{P_{r}(h_{n,f}+ h_{n,b}h_{b,f}\psi_{j})}{\sigma^{2}}, \label{7}
\end{align}
with its corresponding data rate is given as $R_{3} = \frac{1}{2}W\log_{2}(1+\gamma_{3})$. 

Further, if the source and relay nodes are realized by two independent codebooks, then the channels of $BS-U_f$ through BST in the first time slot, $U_n-U_f$ plus $U_n-U_f$ through BST in the second time slot, can be considered as a set of parallel channels. Thus, the resultant maximum rate achieved at the far user can be expressed as follows \cite{mo2014cooperative,maric2010bandwidth}:
\begin{align}
R_{2}+R_{3}=\frac{W}{2}\Big( \log_{2}(1+\gamma_{2}) + \log_{2}(1+\gamma_{3})\Big) 
\end{align}
 Moreover, considering decode-and-forward protocol with $U_n$ successfully decodes the $U_f$ signal, the end-to-end maximum rate achieved at $U_f$ is expressed as $min\{R_{nf}, (R_2+R_3)\}$\cite{maric2010bandwidth, laneman2004cooperative}. However, the near $U_n$ can only decode the $U_f$ signal successfully if the rate of $U_f$ at $U_n$ is greater than or equal to the resultant maximum rate achieved at the $U_f$. Thus, to achieve successful cooperation from $U_n$, we consider that the rate of the far user at near user, $R_{nf}$, must satisfy the following condition as\cite{ali2019joint}:
\begin{align}
	&  \frac{W}{2}\log_{2}(1+\gamma_{nf})\geq \frac{W}{2}\Big(\log_{2}(1+\gamma_{2}) +\log_{2}(1+\gamma_{3})\Big), \label{000000}
\end{align}

Similarly, to fulfill the minimum QoS requirement, we consider that the achievable rates for $U_n$ and $U_f$ must satisfy the following conditions
\begin{align}
	  \frac{W}{2}\log_{2}(1+\gamma_{1})\geq R_{min}, \label{0000001}
\end{align}
\begin{align}
	&  \frac{W}{2}\Big( \log_{2}(1+\gamma_{2}) + \log_{2}(1+\gamma_{3})\Big) \geq R_{min}, \label{0000002}
\end{align}
where $R_{min}$ represents the minimum data rate to meet the QoS requirement of $U_n$ and $U_f$ for the considered system. If $P_{c}$ denotes the circuit power of the system, then the total power consumption $P_{T}$ can be expressed as follows:
\begin{align}
	P_{T}= P\left( \xi_{n}+ \xi_{f}\right)+P_{r}+P_{c}, \label{77}
\end{align}
Next, the main objective of this proposed work is to maximize the EE of the backscatter-enabled cooperative NOMA system under imperfect SIC decoding at the receiver. The total EE of the considered system can be expressed as follows:
\begin{align}
EE=\dfrac{R_{sum}}{P_{T}}, \label{8}
\end{align}
where $R_{sum}$ represents the total sum-rate of BSC cooperative NOMA network and can be expressed as $R_{sum} = R_{1}+ (R_{2}+ R_{3})$ provided that $ (R_{2}+ R_{3})\leq R_{nf}$. Moreover, the EE of considered system can be maximized by efficiently optimizing the transmit power of the source, PAC of NOMA users, and power of relay node. Besides EE maximization, we also desire to consider the QoS, power budgets, and cooperation constraints under the imperfect SIC decoding to acquire the minimum rate requirements for NOMA users. Mathematical expression for a joint EE maximization problem (P) is given as follows:
\begin{alignat}{2}
&\text{(P)}\ \quad \underset{{(P,\xi_{n}, \xi_{f}, P_{r})}}{\text{max}} \dfrac{R_{sum}}{P_{T}} \label{9}\\
s.t.\quad&\text{C1}:\frac{W}{2}\log_{2}(1+\gamma_{1})\geq R_{min}, \nonumber\\&
\text{C2}: 
\frac{W}{2}\Big( \log_{2}(1+\gamma_{2}) + \log_{2}(1+\gamma_{3})\Big) \geq R_{min}, \nonumber\\&
\text{C3}:
   \frac{W}{2}\log_{2}(1+\gamma_{nf})\geq \frac{W}{2}\Big(\log_{2}(1+\gamma_{2}) +\log_{2}(1+\gamma_{3})\Big),  \nonumber\\&  
\text{C4:}\ 0\leq P\left( \xi_{n}+ \xi_{f}\right) \leq P_{t}, \nonumber\\
&\text{C5:} \ 0\leq P_{r} \leq P_{r(max)} ,\nonumber\\
&\text{C6:} \ \xi_{n}+\xi_{f}\leq 1.\nonumber
\end{alignat}
where $P_{t}$ and $P_{r(max)}$ denote the total power budget of source and relay node, respectively. Constraints C1, C2, and C3, obtained from (9), (10), and (11), ensure the successful cooperation and QoS demand for NOMA users. Constraints C4 and C5 limit the transmit power based on the total power budget of the source and relay node, respectively. Finally, constraint C6 restricts the PAC values of $U_{n}$ and $U_{f}$ within the practical range.

Further, the optimization problem defined by (\ref{9}) has non-linear fractional form and it is very difficult to solve. Therefore, we utilize successive convex approximation (SCA) \cite{papandriopoulos2009scale} which transforms the considered problem into a tractable concave-convex fractional programming (CCFP) problem with low complexity. The SCA approximation gives the following lower bound: 
\begin{alignat}{2}
\Phi \log_{2}(\gamma)+\Psi \leq \log_{2}(1+\gamma),  \label{555} 
\end{alignat} 
where $\Phi=\frac{\gamma_{o}}{1+\gamma_{o}}$, $\Psi=\log_{2}(1+\gamma_{o})-\frac{\gamma_{o}}{1+\gamma_{o}}\log_{2}(\gamma_{o})$ are the approximation constants and the bound becomes tight at $\gamma=\gamma_{o}$.

Based on the SCA approximation given in (\ref{555}), the sum-rate of considered system can be expressed as follows: 
\begin{alignat}{2}
 \tilde{R}_{sum}=\sum_{n=1}^{3}\frac{1}{2}(\Phi_{n} \log_{2}(\gamma_{n})+\Psi_{n}),  \label{999} 
\end{alignat} 
where
\begin{alignat}{2}
	\Phi_{n}=\frac{\gamma_{n_o}}{1+\gamma_{n_o}},  \label{9999} 
\end{alignat}
and
\begin{alignat}{2}
\Psi_{n}=\log_{2}(1+\gamma_{n_o})-\frac{\gamma_{n_o}}{1+\gamma_{n_o}}\log_{2}(\gamma_{n_o}),  \label{99999} 
\end{alignat}
Thus, the updated EE maximization problem in (\ref{9}) can be reformulated as follows:
\begin{alignat}{2}
	&\text{(P1)}\ \quad \underset{{(P,\xi_{n}, \xi_{f}, P_{r})}}{\text{max}} \dfrac{\tilde{R}_{sum}}{P_{T}} \label{99}\\
	s.t.\quad&\text{C1}: \gamma_{1} \geq\left(2^{\frac{2R_{min}-\Psi_{1}W}{\Phi_{1}W}}\right)\nonumber\\  &
	\text{C2}: 
	\gamma_{2}^{\Phi_{2}}+ 	\gamma_{3}^{\Phi_{3}}
	  \geq \left (2^{\frac{2R_{min}-\Psi_{2}W-\Psi_{3}W}{W}}\right) ,   \nonumber\\&
	\text{C3}:
 2^{(\Psi_{nf}-\Psi_{2}-\Psi_{3})} \geq \gamma_{2}^{\Phi_{2}}+ 	\gamma_{3}^{\Phi_{3}}-\gamma_{nf}^{\Phi_{nf}}, \nonumber\\&  
	\text{C4:}\ 0\leq P\left( \xi_{n}+ \xi_{f}\right) \leq P_{t}, \nonumber\\
	&\text{C5:} \ 0\leq P_{r} \leq P_{r(max)} ,\nonumber\\
	&\text{C6:} \ \xi_{n}+\xi_{f}\leq 1.\nonumber
\end{alignat}

\section{Energy Efficiency Maximization Solution}
The concave-convex fractional optimization problem given in (\ref{99}) is a non-convex problem. Thus, it is very difficult to find a global optimal solution due to the coupled variables i.e., $P$, $\xi_{n}$, $\xi_{f}$, and $P_{r}$. Therefore, a sub-optimal solution could be found by using an alternating optimization algorithm which solves the problem in three stages as: i) In first stage, transmit power $P$ is computed for the fixed values of $\xi_{n}$, $\xi_{f}$, and $P_{r}$; ii) The PAC, $\xi_{n}$, $\xi_{f}$, are computed in second stage for given values of $P^*$ and $P_{r}$; iii) Finally, in third stage, relay power $P_{r}$ is computed for given $P^*$, $\xi^*_{n}$, and $\xi^*_{f}$.


\subsection{Efficient Transmit Power Allocation for BS}
The EE maximization problem given in (\ref{99}) can be simplified to an efficient power allocation problem for the given values of PAC and $P_{r}$. Thus, the subproblem to optimize the transmit power of BS can be expressed as follows: 
\begin{alignat}{2}
	& \underset{{P}}{\text{max}} {EE}= \underset{{P}}{\text{max}} \frac{\tilde{R}_{sum}}{P_T},\label{12} \\
	& s.t.\quad \text{C1}-\text{C4} \nonumber.
\end{alignat}
The objective function in (\ref{12}) is still non-convex. Thus, we exploit the parameter transformation based on the Dinkelbach method in order to reduce the complexity of the solution. Consider the maximum energy efficiency of the system given as follows  \cite{dinkelbach1967nonlinear}:
\begin{alignat}{2}
	\gamma^*_{EE}= \underset{{P}}{\text{max}} \frac{\tilde{R}_{sum}}{P_T},\label{112} 
\end{alignat}  
Note that the problem in (\ref{112}) is a non-linear concave-convex fractional programming problem which can be transformed into an equal parameterized non-fractional subtractive form given as \cite{dinkelbach1967nonlinear,fang2017joint}:
\begin{alignat}{2}
	&\underset{{P}}{\text{max}} \ \tilde{R}_{sum}-\gamma_{EE}P_{T},\label{18}\\
	& s.t.\quad \text{C1}-\text{C4}.\nonumber 
\end{alignat}
where $\gamma_{EE}$ denotes a scaling parameter for $P_{T}$. Consider a function given as: 
\begin{alignat}{2}
	F(\gamma_{EE})=\underset{{P}}{\text{max}}  \{\tilde{R}_{sum}-\gamma_{EE}P_{T}\},\label{118}
\end{alignat}
Note that $F(\gamma_{EE})$ results in a negative quantity when $\gamma_{EE}$ approaches to $\infty$ and positive quantity while $\gamma_{EE}$ approaches to $-\infty$. Hence, $F(\gamma_{EE})$ is an affine function with respect to $\gamma_{EE}$ \cite{fang2017joint}. Consequently, solving the optimization problem in (\ref{12}) is analogous to determine the maximum energy efficiency $\gamma^*_{EE}$. Further, $\gamma^*_{EE}$ can be achieved if and only if \cite{fang2017joint}:  
\begin{alignat}{2}
	F(\gamma^*_{EE})=\underset{{P}}{\text{max}}  \{\tilde{R}_{sum}-\gamma^*_{EE}P_{T}\}=0,\label{1118}
\end{alignat}

Next, based on the SCA approximation, the sum-rate in (\ref{18}) can be expressed as follows :
	\begin{alignat}{2}
	&\tilde{R}_{sum}= 
	\frac{W}{2}\bigg\{\bigg(\Phi_{1}\log_2\bigg(\frac{P\theta}{P\Theta+n}\bigg)+ \Psi_{1}\bigg)+ \bigg(\Phi_{2}\log_2\bigg(\frac{P\Gamma}{P\Delta+n}\bigg) + \Psi_{2}\bigg) 
	 +\bigg(\Phi_{3}\log_2\bigg(\frac{P_{r}\varphi}{n}\bigg)+ \Psi_{3}\bigg\},\label{17} 
\end{alignat}
where $\theta= \xi_{n}(g_{s,n}+g_{s,b}g_{b,n}\psi_{i})$, $\Theta= \xi_{f}\eta(g_{s,n}+g_{s,b}g_{b,n}\psi_{i})$, and $\sigma^{2}=n$, $\Gamma=\xi_{f}(g_{s,b}h_{b,f}\psi_{i})$, $\Delta=\xi_{n}(g_{s,b}h_{b,f}\psi_{i})$, and $\varphi=h_{n,f}+ h_{n,b}h_{b,f}\psi_{j}$.
	
Next, we exploit the Lagrange dual method and sub-gradient method to compute the sub-optimal solution of optimization problem defined in (\ref{18}). The Lagrangian function of considered optimization problem in (\ref{18}) can be formulated as follows: 
\begin{alignat}{2}
	& L(P,\boldsymbol{\lambda})=	\frac{W}{2}\bigg\{\bigg(\Phi_{1}\log_2\bigg(\frac{P\theta}{P\Theta+n}\bigg)+ \Psi_{1}\bigg)
	 + \bigg(\Phi_{2}\log_2\bigg(\frac{P\Gamma}{P\Delta+n}\bigg) + \Psi_{2}\bigg) 
	 +\bigg(\Phi_{3}\log_2\bigg(\frac{P_{r}\varphi}{n}\bigg)+ \Psi_{3}\bigg\}\nonumber\\
	&-\gamma_{EE}\bigg(P(\xi_{n}+\xi_{f})+ P_{r}+P_{c}\bigg)+ \lambda_{1}\bigg\{P\theta- (2^{\frac{2R_{min}-\Psi_{1}W}{\Phi_{1}W}})
	 (P\Theta + n) \bigg\} +  \lambda_{2}\bigg\{P\Gamma- (2^{\frac{2R_{min}-2\omega-\Psi_{2}W}{\Phi_{2}W}}) (P\Delta + n)\bigg\}\nonumber  \\
	& + \lambda_{3}\bigg\{\bigg( \Phi_{nf}\log_2\bigg(\frac{P\epsilon}{P\delta+n}\bigg)-  \Phi_{2}\log_2\bigg(\frac{P\Gamma}{P\Delta+n}\bigg) \bigg)-
	\bigg( 2\omega-\Psi_{2}-\Psi_{nf} \bigg) \bigg\} +\lambda_{4} \bigg \{P_{t}-P\varPi\bigg\},\label{19}
\end{alignat}
where $\boldsymbol {\lambda}=\{\lambda_{1},\lambda_{2},\lambda_{3},\lambda_{4}\}$ are the Lagrange multipliers for the considered optimization subproblem, $\omega=\frac{1}{2}(\Phi_{3}\log_2(\frac{P_{r}\varphi}{n})+\Psi_{3})$, $\epsilon=\xi_{f}(g_{s,n}+g_{s,b}g_{b,n}\psi_{i})$, $\delta=\xi_{n}(g_{s,n}+g_{s,b}g_{b,n}\psi_{i})$, and $\varPi=\xi_{n}+\xi_{f}$. 

Further, the KKT conditions have been exploited to find the sub-optimal solution in terms of the efficient transmit power allocation for BS. Based on the KKT method, we can write as follows:
\begin{alignat}{2}
\frac{\partial L (P,\boldsymbol{\lambda})}{\partial P}|_{P=P^*}=0, \label{20} 
\end{alignat}
After taking the partial derivative of Lagrangian function $L (P,\boldsymbol{\lambda})$ in (\ref{20}) with respect to $P$, the above equation can be written as follows:
\begin{align}
	& \frac {n\Phi_{1}}{P^{2}\phi+P\tau}+\frac {n\Phi_{2}}{P^{2}\phi_{1}+P\tau}+\frac {P\alpha \lambda_{3}+\beta\lambda_{3}}{P^{3}N_{1}+P^{2}N_{2}+PN_{3}}\nonumber\\
	&+\zeta= 0 \label{21}
\end{align}
where $\phi=\Theta 2ln(2)$, $\tau=n2ln(2)$, $\phi_{1}=\Delta2ln(2)$, $\alpha=n\Delta\Phi_{4}-n\delta\Phi_{2}$, $\beta=n^{2}\Phi_{4}-n^{2}\Phi_{2}$, $N_{1}=2ln(2)$, $N_{2}=n2ln(2)\delta +n2ln(2)\Delta$, and $N_{3}=n^{2}2ln(2)$.
After applying some mathematical computations on Eq. (\ref{21}), we obtain the following quintic polynomial given as follows:
\begin{align}
P^{5}\Lambda_{5}+ P^{4}\Lambda_{4}+ P^{3}\Lambda_{3}+ P^{2}\Lambda_{2}+ P\Lambda_{1}+ \Lambda=0,\label{22}
\end{align}
where $\Lambda_{5}=N_{1}\phi\phi_{1}\zeta$ and $\Lambda=(N_{3}n\Phi_{1}\tau +N_{3}n\Phi_{2}\tau+ \tau^{2}\beta\lambda_{3})$. The values of $\Lambda_{1}$, $\Lambda_{2}$, $\Lambda_{3}$, and $\Lambda_{4}$, and $\zeta$ are given in Eqs. (30), (31), (32), (33), and (34), respectively. Moreover, the optimal solution in terms of $P$ can be easily computed by employing the built-in functions provided in MATLAB and Mathematica solvers. 
\begin{figure*}
	\begin{align}
		\Lambda_{1} &= N_{3}n\Phi_{1}\phi_{1} +N_{3}n\Phi_{2}\phi +N_{2}n\Phi_{1}\tau +N_{2}n\Phi_{2}\tau +\tau^{2}\alpha\lambda_{3} +\tau\phi_{1}\beta\lambda_{3}+ \tau\phi\beta\lambda_{3}+N_{3}\tau^{2}\zeta,\\
		\Lambda_{2}&=  N_{2}n\Phi_{1}\phi_{1} + N_{2}n\Phi_{2}\phi + N_{1}n\Phi_{1}\tau +  N_{1}n\Phi_{2}\tau+ \tau\phi_{1}\alpha\lambda_{3}+ \tau\phi\alpha\lambda_{3}+ \phi\phi_{1}\beta\lambda_{3}+ N_{3}\tau\phi_{1}\zeta \nonumber  \\ &+N_{3}\tau\phi\zeta+ N_{2}\tau^{2}\zeta, \\
		\Lambda_{3}& =N_{3}n\Phi_{1}\phi_{1}+ N_{3}n\Phi_{2}\phi+ \phi \phi_{1}\alpha\lambda_{3}+ N_{3}\phi\phi_{1}\zeta+ N_{1}\tau^{2}\zeta,\\
		\Lambda_{4}&=N_{2}\phi\phi_{1}\zeta +N_{1}\tau\phi_{1}\zeta +N_{1}\tau\phi\zeta,\\
		\zeta&=\lambda_{1}(\theta-\Theta(2^{\frac{2R_{min}-\Psi_{1}}{\Phi_{1}}}))+\lambda_{2}(\Gamma-\Delta(2^{\frac{2R_{min}-2\omega-\Psi_{2}}{\Phi_{2}}}))-\lambda_{4}\varPi-\gamma_{EE}\varPi.     
	\end{align}\hrulefill
\end{figure*}

Subsequently, the Lagrange multipliers $\lambda_{1}$, $\lambda_{2}$, $\lambda_{3}$, and $\lambda_{4}$ are updated in an iterative manner by using sub-gradient method as \cite{khan2019joint}:
\begin{align}
	&\lambda_{1}(l+1)=\lambda_{1}(l)+\mu(l)\times  
	\bigg(( \xi_{n}P\left( g_{s,n}+g_{s,b}g_{b,n}\psi_{i}\right) -\left(2^{\frac{2R_{min}-\Psi_{1}W}{\Phi_{1}W}}\right) \times   \left( \xi_{f}P\eta\left( g_{s,n}+g_{s,b}g_{b,n}\psi_{i}\right) +\sigma^2\right)\bigg ),\label{27}  
\end{align}
\begin{align}
	&\lambda_{2}(l+1)= \lambda_{2}(l)+\mu(l)\times\bigg ((\left( P\xi_{f}g_{s,b}h_{b,f}\psi_{i}\right)^{\Phi_2}\times (\sigma^{2})^{\Phi_{3}}) +  \left(P_{r}\left( h_{n,f}+ h_{n,b}h_{b,f}\psi_{j}\right)\right)^{\Phi_{3}} \nonumber\\ &\times \left( \left( P\xi_{n}g_{s,b}h_{b,f}\psi_{i}\right) + \sigma^{2}\right)^{\Phi_2} 
	-  (2^{\frac{(2R_{min}-\Psi_{2}W-\Psi_{3}W)}{W}}) \times \left( \left(\sigma^{2}\right)^{\Phi_{3}} \times \left( \left( P\xi_{n}g_{s,b}h_{b,f}\psi_{i}\right) + \sigma^{2}\right)^{\Phi_2}\right)\bigg),\label{28}  
\end{align}
\begin{align}
	&\lambda_{3}(l+1)=\lambda_{3}(l)+\mu(l)\times\bigg (\left( 2^{(2R_{nf}-\Psi_{2}-\Psi_{3})}\right) \times  \left( \left(\sigma^{2}\right)^{\Phi_{3}} \times \left( \left( P\xi_{n}g_{s,b}h_{b,f}\psi_{i}\right) + \sigma^{2}\right)^{\Phi_2}\right)   \nonumber \\& 
	- (\left( P\xi_{f}g_{s,b}h_{b,f}\psi_{i}\right)^{\Phi_2}\times (\sigma^{2})^{\Phi_{3}}) + \left(P_{r}\left( h_{n,f}+ h_{n,b}h_{b,f}\psi_{j}\right)\right)^{\Phi_{3}} 
	\times \left( \left( P\xi_{n}g_{s,b}h_{b,f}\psi_{i}\right) + \sigma^{2}\right)^{\Phi_2}\bigg ),\label{29}  
\end{align}
\begin{align}
	\lambda_{4}(l+1)&=\lambda_{4}(l)+\mu(l)\times\bigg(P_{t}-P\left( \xi_{n}+ \xi_{f}\right)\bigg ),\label{30}  
\end{align}
where $l$ is the iteration index and $\mu(l)$ denotes the step size of the sub-gradient method. Note that an appropriate step size is required for the convergence of algorithm.

\subsection{Optimization for Power Allocation Coefficients}
We compute the power allocation coefficients $\xi_{n}$ and $\xi_{f}$ for considered BSC-enabled cooperative NOMA system. For the given values of $P^*$ and $P_{r}$, the EE maximization optimization problem given in (\ref{99}) can be expressed as follows:
\begin{alignat}{2}
	& \underset{{(\xi_{n}, \xi_{f})}}{\text{max}} 
	\frac{\tilde{R}_{sum}}{P_T}= \underset{{(\xi_{n}, \xi_{f})}}{\text{max}}\tilde{R}_{sum}-\gamma_{EE}P_{T},\nonumber \\
	&s.t.\quad \text{C1}-\text{C4}, \text{C6}.\label{31} 
\end{alignat}

Since, the sum-rate in  (\ref{31}) can be expressed as follows
	\begin{alignat}{2}
	&\tilde{R}_{sum}= 
	\frac{W}{2}\bigg\{\bigg(\Phi_{1}\log_2\bigg(\frac{\xi_{n}\theta_{1}}{\xi_{f}\Theta_{1}+n}\bigg)+\Psi_{1}\bigg)
	 +\bigg(\Phi_{2}\log_2\bigg(\frac{\xi_{f}\Gamma_{1}}{\xi_{n}\Gamma_{1}+n}\bigg)+\Psi_{2}\bigg) 
	& +\bigg(\Phi_{3}\log_2\bigg(\frac{P_{r}\varphi}{n}\bigg)+\Psi_{3}\bigg)\bigg\},\label{32}
\end{alignat}
where $\theta_{1}=P(g_{s,n}+g_{s,b}g_{b,n}\psi_{i})$, $\Theta_{1}=P\eta(g_{s,n}+g_{s,b}g_{b,n}\psi_{i})$, $\Gamma_{1}=P(g_{s,b}h_{b,f}\psi_{i})$. 

As, the objective function of non-fractional optimization problem in  (\ref{31}) is an affine function with respect to $\gamma_{EE}$ \cite{fang2017joint}. Thus, we exploit the KKT condition to find the sub-optimal solution of power allocation coefficients. However, for the sake of simplicity, we omit the similar derivation steps adopted in the transmit power allocation subproblem. Consequently, by employing KKT conditions and updating the corresponding Lagrange multipliers using the sub-gradient method, the optimal values of $\xi_{n}$ and $\xi_{f}$ can be computed as follows: 
\begin{align}
 \xi_{n}^{4}\pi_{4}+ \xi_{n}^{3}\pi_{3}+ \xi_{n}^{2}\pi_{2}+ \xi_{n}\pi_{1}+ \pi = 0,\label{33}
\end{align}
\begin{align}
\xi^*_{f}=1-\xi^*_{n} ,\label{34}
\end{align}
where $\pi=N_{4}\Phi_{1}\tau$, $\pi_{4}=\phi_{2}\phi_{3}N_{1}\zeta_{1}$, $\phi_{2}=\Gamma_{1}2ln(2)$, $\phi_{3}=\Gamma_{1}\theta_{1}(2ln(2))^{2}$, $\phi_{4}=(\Gamma_{1}n(2ln(2))^{2}+\theta_{1}n(2ln(2))^{2})$, $N_{4}=(n2ln(2))^{2}$, $\alpha_{1}=\Gamma_{1}\theta_{1}2ln(2)(\Phi_{2}-\Phi_{nf})$, $\beta_{1}=n2ln(2)(\Phi_{2}\Gamma_{1}-\Phi_{nf}\theta_{1})$. The values of $\pi_{1}$, $\pi_{2}$, and $\pi_{3}$, and $\zeta_{1}$ are given in equations (43), (44), (45), and, (46), respectively, where  $\tilde{\gamma}_{1}$, $\tilde{\gamma}_{2}$, $\tilde{\gamma}_{3}$, $\tilde{\gamma}_{4}$, and $\tilde{\gamma}_{5}$ are the Lagrange multiplier associated with constraints C1, C2, C3, C4, and C6, respectively. Moreover, the quartic polynomial given in (\ref{33}) can be easily solved using any conventional solver to find the optimal value of $\xi^*_{n}$. Finally, once the value of $\xi^*_{n}$ is in hand, $\xi^*_{f}$ can be computed using Eq. (\ref{34}). 
\begin{figure*}
	\begin{align}
	\pi_{1} &= N_{1}N_{4}\tau \zeta_{1}+N_{4}\Phi_{1}\phi_{2}+ \Phi_{1}\phi_{4}\tau +N_{1}\tau\beta_{1}\tilde{\gamma}_{3}- N_{1}N_{4}\Phi_{2}\Gamma_{1},\\
	\pi_{2}&= N_{1}N_{4}\phi_{2}\zeta_{1}+ N_{1}\phi_{4}\tau\zeta_{1}+\Phi_{1}\phi_{2}\phi_{4}+ \Phi_{1}\phi_{3}\tau+ N_{1}\tau\alpha_{1}\tilde{\gamma}_{3}+ N_{1}\phi_{2}\beta_{1}\tilde{\gamma}_{3}- N_{1}\Phi_{2}\Gamma_{1}\phi_{4}, \\
	\pi_{3}& =\phi_{2}\phi_{4}N_{1}\zeta_{1}+\phi_{3}\tau N_{1}\zeta_{1}+ \Phi_{1}\phi_{2}\phi_{3}+N_{1}\phi_{2}\alpha_{1}\tilde{\gamma}_{3}+N_{1}\Phi_{2}\Gamma_{1}\phi_{3},\\ 
	\zeta_{1}& = \tilde{\gamma}_{1}\theta_{1}-\tilde{\gamma}_{2}(\Gamma_{1}(2^{\frac{2R_{min}-2\omega-\Psi_{2}}{\Phi_{2}}}))-\tilde{\gamma}_{4}P-\tilde{\gamma}_{5}-\gamma_{EE}P.
	\end{align}\hrulefill
\end{figure*}

\subsection{Efficient Power Optimization for Relay Node}
In this subsection, we compute the optimal transmit power for relay node, $P_{r}$, in the second time slot of the considered BSC-enabled cooperative NOMA system. For the given values of $P^*$, $\xi^*_{n}$, and $\xi^*_{f}$, the EE maximization optimization problem given in Eq. (\ref{99}) can be written as follows:
\begin{alignat}{2}
& \underset{{P_{r}}}{\text{max}} 
\frac{\tilde{R}_{sum}}{P_T}= \underset{{P_{r}}}{\text{max}} \tilde{R}_{sum}-\gamma_{EE}P_{T},\nonumber \\
	&s.t.\quad \text{C2}, \text{C3}, \text{C5}.\label{40} 
\end{alignat}
Next, the sum-rate given in (\ref{40}) can be written as follows: 
	\begin{alignat}{2}
		&\tilde{R}_{sum}= 
		\frac{W}{2}\bigg\{\bigg(\Phi_{1}\log_2\bigg(\frac{\xi_{n}\theta_{1}}{\xi_{f}\Theta_{1}+n}\bigg)+\Psi_{1}\bigg)
	 +\bigg(\Phi_{2}\log_2\bigg(\frac{\xi_{f}\Gamma_{1}}{\xi_{n}\Gamma_{1}+n}\bigg)+\Psi_{2}\bigg) 
	 +\bigg(\Phi_{3}\log_2\bigg(\frac{P_{r}\varphi}{n}\bigg)+\Psi_{3}\bigg)\bigg\},\label{41}
	\end{alignat}

 Next, KKT conditions can be exploited to compute the optimal value of $P_{r}$. The Lagrangian function of considered optimization problem in (\ref{40}) can be formulated as follows:
\begin{alignat}{2}
	& L(P_{r},\boldsymbol{\Upsilon})=		\frac{1}{2}\bigg\{\bigg(\Phi_{1}\log_2\bigg(\frac{\xi_{n}\theta_{1}}{\xi_{f}\Theta_{1}+n}\bigg)+\Psi_{1}\bigg)
	 +\bigg(\Phi_{2}\log_2\bigg(\frac{\xi_{f}\Gamma_{1}}{\xi_{n}\Gamma_{1}+n}\bigg)+\Psi_{2}\bigg) 
	 +\bigg(\Phi_{3}\log_2\bigg(\frac{P_{r}\varphi}{n}\bigg)+\Psi_{3}\bigg)\bigg\}\nonumber\\
	&-\gamma_{EE}\bigg(P(\xi_{n}+\xi_{f})+ P_{r}+P_{c}\bigg)+ \Upsilon_{1}\bigg\{P_{r}\varphi 
	 -n\bigg(2^{\frac{2R_{min}-2\omega_{f}-\Psi_{3}W}{\Phi_{3}W}}\bigg)\bigg\} +  \Upsilon_{2}\bigg\{n\bigg(2^{\frac{2\omega_{n,f}-2\omega_{f}-\Psi_{3}W}{\Phi_{3}W}}\bigg)-P_{r}\varphi\bigg\}\nonumber  \\
	& + \Upsilon_{3}\bigg\{P_{r(max)}-P_{r}\bigg\} ,\label{42}
\end{alignat}
where $\boldsymbol {\Upsilon}=\{\Upsilon_{1},\Upsilon_{2}, \Upsilon_{3}\}$ are the Lagrange multipliers, $\omega_{f}=\frac{1}{2}(\Phi_{2}\log_2(\frac{\xi_{f}\Gamma_{1}}{\xi_{n}\Gamma_{1}+n})+\Psi_{2})$, and $\omega_{n,f}=\frac{1}{2}(\Phi_{nf}\log_2(\frac{\xi_{f}\theta_{1}}{\xi_{n}\theta_{1}+n})+\Psi_{nf})$. 
Next, based on KKT conditions, we can write as follows:
\begin{alignat}{2}
	\frac{\partial L (P_{r},\boldsymbol{\Upsilon})}{\partial P_{r}}|_{P_{r}=P^*_{r}}=\frac{\Phi_{3}}{P_{r}N_{1}}-\zeta_{2} =0, \label{43} 
\end{alignat}
where $\zeta_{2}=(\Upsilon_{3}+\Upsilon_{2}\varphi+\gamma_{EE}-\Upsilon_{1}\varphi)$. After some mathematical computation, the closed-form expression for the optimal transmit power of relay node can be obtained as follows:
\begin{alignat}{2}
P^*_{r}=\bigg[\frac{\Phi_{3}}{\zeta_{2}N_{1}}\bigg]^+, \label{44} 
\end{alignat}
where $[x]^+=max[0,x]$.
\subsection{Proposed Algorithm and Complexity Analysis}
In this subsection, we propose an energy efficiency maximization algorithm for the proposed BSC cooperative NOMA system in terms of the optimal solutions of transmit power $P$, PAC $\xi_{n}$, $\xi_{f}$, and relay power $P_{r}$ provided in Sections III-A, III-B, and III-C, respectively. First, the optimal transmit power $P^*$ is computed for given values of $\xi_{n}$, $\xi_{f}$, and $P_{r}$. Next, with $P^*$ in hand, the optimal values of PAC $\xi^*_{n}$, $\xi^*_{f}$ are computed. Finally, for given values of $P^*$, $\xi^*_{n}$, $\xi^*_{f}$, the optimal relay power $P^*_{r}$ is computed. Moreover, the Lagrange dual variables are updated iteratively until convergence is achieved. Note that the theoretical analysis in terms of the convergence of proposed algorithm is not provided, because the convergence is achieved by exploiting the Dinkelback transformation \cite{dinkelbach1967nonlinear}. Since, all of the sub-problems involve in this work are CCFP problems, and it is already analyzed proved by Dinkelbach that it has linear convergence for CCFP problems. However, we have analyzed the numerical convergence of proposed algorithm based on Dinkelbach transformation (please refer to Fig.\ref{f6} in Section V). Further, let $\mathcal J_1$, $\mathcal J_2$, and $\mathcal J_3$ denote the number of iterations required for Stage 1, Stage 2, and Stage 3, respectively. If, $\mathcal I$ represents the number of iterations required for the convergence of overall algorithm, then the complexity of \textbf{Algorithm 1} can be computed as $O[\mathcal I(\mathcal J_1+\mathcal J_2+\mathcal J_3)]$.         
  \begin{algorithm}[!t]
	{\bf Initialization:} Initialize step sizes, iteration index, $j=j_{1}$, $j_{2}$, $j_{3}=1$, and dual variables  \\
	\While{not converge}{
	 \textbf{Stage 1:} Compute transmit power $P$, for the fixed values of PAC and $P_r$.\\
		\While{not converge}{
	\For{$j_1=1:\mathcal J_1$}{Compute $\gamma_{EE}(j_{1})$=$\frac{R_{sum}(j_{1})}{P_T}$\\ 
		Update $\lambda_{1}, \lambda_{2}, \lambda_{3}, \lambda_{4}$ using  Eqs. (\ref{27})-(\ref{30})\\
			Compute optimal transmit power $P^*$ using Eq. (\ref{22})\\}
		}
	 \textbf{Stage 2:} With optimal $P^*$ and fixed $P_r$, compute PAC\\
\While{not converge}{
	\For{$j_2=1:\mathcal J_2$}{Compute $\gamma_{EE}(j_{2})$\\ 
		Update Lagrange dual varibales\\
		Compute optimal value of $\xi^*_{n}$ using Eq. (\ref{33}) \\
    	Compute optimal value of $\xi^*_{f}$ using Eq. (\ref{34}) \\ }
}

	 \textbf{Stage 3:} With optimal $P^*$ and PAC in hand, compute $P_r$\\
\While{not converge}{
	\For{$j_3=1:\mathcal J_3$}{Compute $\gamma_{EE}(j_{3})$\\ 
		Update Lagrange dual varibales\\
		Compute the optimal $P^*_{r}$ using Eq. (\ref{44}) \\ }
}

}
	Return $P^*$, $\xi^*_{n}$, $\xi^*_{f}$, $P^*_{r}$
	\caption{Proposed Alternating Optimization Algorithm for considered BST-Assisted Cooperative NOMA}
\end{algorithm}  
   \section{Jointly-Designed QC-LDPC Codes for BSC Cooperative NOMA}
In cooperative NOMA, the relay node exploits the SIC principle to decode the data for a distant paired user. However, if the data for a far user is decoded with an error, then this error will propagate and the performance of both relay and destination node would be affected. Hence, to avoid this chain process of error propagation, efficient error-correction codes could be utilized at relay and destination nodes. Therefore, the central focus of this Section is to provide jointly-designed QC-LDPC codes for considered BSC-enabled cooperative NOMA network. The far user $U_{f}$ suffers from performance degradation in terms of bit-error probability due to bad channel conditions. To enhance the BER performance of $U_{f}$, we propose a joint framework for QC-LDPC codes to jointly decode $U_{f}$ data received from the source (backscatter-aided signal) and $U_{n}$ in $1^{st}$ and $2^{nd}$ time frames, respectively. More specifically, at the source node, two parity-check matrices $\mathbf{H}^{(1)}$ and $\mathbf{H}^{(2)}$ would be utilized to encode the $U_{n}$ and $U_{f}$ signals, respectively, before the NOMA-enabled SC coding, where the relay node is realized by a parity-check matrix $\mathbf{H}^{(3)}$ to decode the incoming superimposed signal from the BS. During the first time slot, the source node broadcasts the encoded superimposed data towards $U_{n}$ and $U_{f}$. The near user $U_{n}$ directly receives data from BS, but, the far user $U_{f}$ only receives signal reflected by the backscatter due to the obstruction between BS and $U_{f}$. In the second time frame, the relay node first decodes the far user data, then broadcasts it towards backscatter and far user $U_{f}$. In the destination node, the signals are alternatively received by a matched filter coming from the backscatter-aided source and $U_{n}$ working as a relay node. The received signals are multiplexed, and finally decoded by SPA-based joint iterative decoder \cite{asif2020jointly}, where decoder utilizes a parity-check matrix $\mathbf{H}$, a joint combination of $\mathbf{H}^{(1)}$, $\mathbf{H}^{(2)}$, and $\mathbf{H}^{(3)}$, to jointly decode the $U_{f}$ signal. 
  
\subsection{Basic Preliminaries}
 In this Section, we first provide some basic preliminaries and fundamental concepts for balanced incomplete block design (BIBD)  \cite{colbourn2007crc}, \cite{zhang2007spectrum} and CBSEC to facilitate the readability of readers. 
 
 ${\textbf{Definition 1 }:}$ A design can be represented by a pair $(\mathcal R,\mathcal B)$, where $\mathcal R$ is total number of varieties hold by a set, and $\mathcal B$ consists of subsets (non-empty), also called blocks, of $\mathcal R$. For any  three positive integers $\Omega$, $f$, and $\chi$, suppose that $\Omega>f\geq2$. A $(\Omega, f, \chi)$-BIBD, defined a pair $(\mathcal R,\mathcal B)$, must hold the properties given below:
 \begin{enumerate}
 	\setlength{\itemsep}{1.7ex}
 	\item[(1)] $|\mathcal R|=\Omega$,
 	\item[(2)] there are maximum $f$ number of varieties hold by each subset of $\mathcal B$, and
 	\item[(3)] if there are exactly $\chi$ blocks in which each pair of the elements appears.
 \end{enumerate}
 
 ${\textbf{Definition 2} :}$ A pair $(\mathcal R,\mathcal B)$ is called balanced sampling plans excluding contiguous units with $\mathcal R$ = $Z_{\Omega}$ and $\mathcal B$ denotes the nonempty blocks of $\mathcal R$, such that no elements pair $(x,y)$ appears in any subset of $\mathcal B$ if $x-y$= $\pm1,...,\pm\Xi$ $(mod$ $\Omega)$, however, any other pair of elements participates exactly in $\chi$ subsets, where $\Xi$ denotes a natural number.
  
 Next, the notation $(\Omega, f, \chi;\Xi)$-BSEC will be used to denote a balanced sampling plan excluding contiguous units. Let $Z_{\Omega}$ be a cyclic additive group of order $\Omega$ and $(\mathcal R,\mathcal B)$ a $\Omega, f, \chi;\Xi)$-BSEC. A design$(\mathcal R,\mathcal B)$ is called cyclic if $Z_{\Omega}$ is an automorphism of $(\Omega, f, \chi;\Xi)$-BSEC. We use the notation $(\Omega, f, \chi;\Xi)$-CBSEC to denote a cyclic BSEC \cite{zhang2005spectrum}, \cite{hedayat1988sampling}.
   
 Subsequently, we briefly describe the construction of $(\Omega, f, \chi;\pi)$-CBSEC based on perfect and hooked Langford sequences \cite{simpson1983langford}. Suppose that $A \subseteq Z_{\Omega}$. Let $\Delta A$= $\{a_{i}-a_{j} \:(mod\:\Omega): a_{i},a_{j}\in A, a_{i}\neq a_{j}\}$. Let $[x,y]$ be a set of all integers $r$ such that $x\leq r\leq y$, and $\chi[x,y]$ represents a multi-set which contains each element of $[x,y]$ $\chi$ times. From literature \cite{hedayat1988sampling}, we can obtain the following Lemma given as follows:
  
  $\textbf{Lemma 1}:$ Suppose that there exist $f$-subsets (non-empty) $A_{1},A_{2},...,A_{w}$ of $Z_{\Omega}$ such that the multi-set union $\bigcup_{i=1}^{w} \Delta A_{i}=\chi[\Xi+1,\Omega-\Xi +1]$, then there exists a $(\Omega, f, \chi;\Xi)$-CBSEC.
  
  The nonempty k-subsets $A_{1},A_{2},...,A_{w}$ of $Z_{\Omega}$ are called the base blocks of $(\Omega, f, \chi;\Xi)$-CBSEC. $\textbf{Lemma 1}$ enables us to construct $(\Omega, f, \chi;\Xi)$-CBSEC based on perfect and hooked Langford sequences \cite{simpson1983langford}.

  ${\textbf{Definition 3}:}$ A sequence $E=(e_{1}, e_{2},..., e_{2w})$ of $2w$ elements is called a Langford sequence (order $w$) with defect $\mathcal V$ under all of the properties given as:
  \begin{enumerate}
  	\setlength{\itemsep}{1.7ex}
  	\item[(i)] for each $s\in \{ \mathcal V, \mathcal V+1...,  \mathcal V+ w -1\}$ there exist exactly two elements $e_{i},e_{j}\in E$ such that $e_{i}=e_{j}=s$, and
  	\item[(ii)] if $e_{i}=e_{j}=s$ with $i<j$, then $j-i=s$.
  \end{enumerate}

  ${\textbf{Definition 4}:}$ A sequence of $2w+1$ elements, $E=(e_{1},e_{2}..., e_{2w-1}, 0, e_{2w+1})$,  is called a hooked Langford sequence with order $w$ and defect $\mathcal V$ under all of the properties given as follows:
  \begin{enumerate}
  	\setlength{\itemsep}{1.7ex}
  	\item[(i)] for each $s\in \{\mathcal V,\mathcal V+1..., \mathcal V+ w -1\}$ there exist exactly two elements $e_{i},e_{j}\in E$ such that $e_{i}=e_{j}=s$,
  	\item[(ii)] if $e_{i}=e_{j}=s$ with $i<j$, then $j-i=s$, and
  	\item[(iii)] a zero (hook) is introduced at $2w$ position of set $E$.
  \end{enumerate}
  
  Further, $\textbf{Lemma 2}$ provides the existence of Langford sequences given as follows.
  
 {\textbf{Lemma 2} :} If $w\geq 2\mathcal V-1$, then $[1,2w]$ can be partitioned into a set of differences $\{\mathcal V,\mathcal V+1..., \mathcal V+w -1\}$ whenever $(w,\mathcal V)\equiv$ $(0,1),(1,1),(0,0),(3,0)$ $(mod (4,2))$.
  
  If $w(w-2\mathcal V+1)+2\geq 0$, then $[1,2w+1]\setminus\{2w\}$ can be partitioned into a set of differences $\{\mathcal V,\mathcal V+1..., \mathcal V+w -1\}$ whenever $(w,\mathcal V)\equiv$ $(2,0),(1,0),(2,1),(3,1)$ $(mod (4,2))$.
  
  The following two lemmas enable us to construct $(\Omega, f, \chi;\Xi)$-CBSEC based on Langford sequences given as follows \cite{simpson1983langford}:
  
  {\textbf{Lemma 3}:} Let $(w,\mathcal V)\equiv$ $(0,1),(1,1),(0,0),(3,0)$ $(mod (4,2))$ with $w\geq 2\mathcal V-1$. Thus, the interval $[\mathcal V,\mathcal V+3w-1]$ can be divided into triples or sets as $\{x_{i},y_{i},z_{i}\}$, $1\leq i\leq w$, such that $x_{i}+y_{i}=z_{i}$.
  
  {\textbf{Lemma 4}:} Let$(w, \mathcal V )\equiv$ $(2,0),(1,0),(2,1),(3,1)$ $(mod (4,2))$ such that $w(w-2\mathcal V +1)+2\geq 0$. Then $[\mathcal V , \mathcal V+3w]\setminus\{\mathcal V+3w-1\}$ can be divided into sets $\{x_{i},y_{i},z_{i}\}$, $1\leq i\leq w$, such that $x_{i}+y_{i}=z_{i}$.
  
  Based on above discussion, Langford sequences can be used to construct $(\Omega, f, \chi;\Xi)$-CBSEC of order $\Omega$ using the steps given as follows:
  \begin{enumerate}
  	\setlength{\itemsep}{1.7ex}
  	\item[(i)] build pairs $(i, j)$, where $e_{i}=e_{j} \in E$;
  	\item[(ii)] change each pair $(i, j)$ into sets of triples as $\{x_{i},y_{i},z_{i}\}$ such that $x_{i}+y_{i}=z_{i}$, $1\leq i\leq w$, where $x_{i}=j-i$, $y_{i}=i+\mathcal V+ w-1$, and $z_{i}=j+\mathcal V+ w-1$;
  	\item[(iii)] the base blocks $\{0, x_{i},z_{i}\}$ can be constructed using each triple;
  	\item[(iv)] add $1$ to each base block set $\{0, x_{i},z_{i}\}$ $(mod$ $\Omega)$ to construct a $(\Omega, f, \chi;\Xi)$-CBSEC.
  \end{enumerate}

Next, we construct a class of jointly-designed QC-LDPC codes based on the cyclic BSEC, and structural properties of finite field.  
  \subsection{CBSEC-Based QC-LDPC Codes}
  Based on above discussion, we utilize the $(\Omega, f, \chi;\Xi)$-CBSEC construction to design a length-4 cycles free QC-LDPC codes. Consider a base matrix $\mathbf{E^{(1)}}$ given as follows: 
  \begin{align}
  	\mathbf{E}^{(1)}=\begin{bmatrix}
  		\mathbf{E}_{1}\\
  		\mathbf{E}_{2}
  	\end{bmatrix}=\begin{bmatrix}
  		{\tilde{P}_{1,1}} & 	{\tilde{P}_{1,2}} & \cdots  & 	{\tilde{P}_{1,w}}\\
  		\breve{P}_{2,1} & \breve{P}_{2,2} & \cdots &\breve{P}_{2,w}
  	\end{bmatrix} , \label{45}
  \end{align}
where the base matrix $\mathbf{E}^{(1)}$ consists of two rows blocks $\mathbf{E}_{1}$ and $\mathbf{E}_{2}$. Both of the row blocks contain $f\times f$ circulant permutation matrices ${\tilde{P}_{1,i}}$ and $\breve{P}_{2,i}$, respectively, $1\leq i\leq w$, constructed using $(\Omega, f, \chi;\Xi)$-CBSEC. More specifically, the $f\times f$ circulant matrices ${\tilde{P}_{1,i}}$, $1\leq i\leq w$, appearing in the first row block $\mathbf{E}^{(1)}$, are constructed using the base blocks of $(\Omega, f, \chi;\Xi)$-CBSEC, where the base block denotes the first row of ${\tilde{P}_{1,i}}$, $1\leq i\leq w$, while all of the other rows are obtained from cyclic-shift operation. Moreover, the $f\times f$ circulant matrices $\breve{P}_{2,i}$, $1\leq i\leq w$, participating in second row block are constructed from negative base blocks of $(\Omega, f, \chi;\Xi)$-CBSEC (mod $\Omega$) \cite{colbourn2007crc}. Hence, the base matrix $\mathbf{E}^{(1)}$ in (\ref{45}) gives a length-4 cycles free QC-LDPC code, where the minimum distance of designed code is lower bounded by $2f+1$ and rate is equal to $(w-2)/w$ \cite{ryan2009channel}.  
 
Further, let $GF(\varsigma)$ is a finite field, where $\varsigma$ denotes the total number of varieties (elements) exist in field $GF(\varsigma)$. For each non-zero element $\phi_{11}^{s}$ of $GF(\varsigma)$, we construct a $(\varsigma -1)$-tuple $\boldsymbol \eta_{b}(\phi_{11}^{s})=(\eta_{0},\eta_{1},..., \eta_{\varsigma-2})$, $0\leq s < \varsigma-1$, based on the binary field $GF(\varsigma =2)$, where $\phi_{11}$ is a primitive element of finite field $GF(\varsigma)$. All of the elements of $\boldsymbol \eta_{b}$ are zero except the $sth$ entry $\mu_{s}=1$. Moreover, if $\phi_{11}^{s}=0$, the binary $(\varsigma -1)$-tuple is represented by an all-zero vector as  $\boldsymbol \eta_{b}(\phi_{11}^{s}=0)=(0, 0,..., 0)$.    

Furthermore, for any element $\psi_{11}\phi_{11}$ over $GF(\varsigma)$, the binary $(\varsigma -1)$-tuple of $\psi_{11}\phi_{11}$ over $GF(\varsigma)$ can be obtained from the cyclic-shift of binary $(\varsigma -1)$-tuple $\boldsymbol \eta_{b}(\psi_{11})$ of any field element $\psi_{11}$. Thus, a $(\varsigma -1)\times (\varsigma -1)$ binary circular permutation matrix, $\mathbf{Z}_{b}(\psi_{11})$, can be constructed based on the $(\varsigma -1)$-tuples of $\psi_{11}$, $\psi_{11}\phi_{11}$, $\phi_{11}^{2}\psi_{11}$,..., $\phi_{11}^{q-2}\psi_{11}$. However, the binary matrix dispersion $\mathbf{Z}_{b}(\psi_{11}=0)$ returns a $(\varsigma -1)\times (\varsigma -1)$ all-zero matrix over $GF(\varsigma =2)$ 

Based on above discussion, substituting non-zero elements $\psi$ of base matrix $\mathbf{E}^{(1)}$ in (\ref{45}) by $(\varsigma -1)\times (\varsigma -1)$ binary dispersion matrix $\mathbf{Z}_{b}(\psi_{11})$ and zero-elements by their $(\varsigma -1)\times (\varsigma -1)$ all-zero matrices $\mathbf{Z}_{b}(0)$, We obtain a $2f\times wf$ array $\mathbf{H}_{b}^{(1)}$ given as follows:            
  \begin{align}
  	&\mathbf{H}_{b}^{(1)}=\nonumber \\&
  	\scriptscriptstyle{\begin{bmatrix}
  		\mathbf{Z}(0,0) & \mathbf{Z}(0,1) & \cdots  & \mathbf{Z}(0,fw-1)\\
  		\mathbf{Z}(1,0) & \mathbf{Z}(1,1) & \cdots  & \mathbf{Z}(1,kw-1)\\
  		\vdots & \vdots  & \ddots & \vdots  \\
  		\mathbf{Z}(2f-1,0) & \mathbf{Z}(2f-1,1) & \cdots  & \mathbf{Z}(2f-1,wf-1) 
  	\end{bmatrix}}, \label{46}
  \end{align} 
where each $\mathbf{Z}(i,j)$ in (\ref{46}) denotes a $(\varsigma -1)\times (\varsigma -1)$ matrix dispersion over binary field $GF(\varsigma =2)$, for $0\leq i < 2f$, $0\leq j < wf$. Thus, the array $\mathbf{H}_{b}^{(1)}$ returns a $2f(\varsigma -1)\times wf(\varsigma -1)$ binary matrix over $GF(\varsigma =2)$. Moreover, the binary matrix $\mathbf{H}_{b}^{(1)}$ also satisfies the RC-constraint \cite{ryan2009channel}. Thus, the null space of $\mathbf{H}_{b}^{(1)}$ gives a class of binary length-4 cycles free QC-LDPC codes.

Suppose that $\lambda_{c}$ and $\mu_{r}$ are two positive integers, for $1\leq \lambda_{c}\leq 2f$ and $1\leq \mu_{r}\leq wf$. Let $\mathbf{H}_{b}^{(1)}(\lambda_{c}, \mu_{r})$ is a $\lambda_{c}\times \mu_{r}$ subarray of $\mathbf{H}_{b}^{(1)}$ and gives a $\lambda_{c}(\varsigma -1)\times \mu_{r}(\varsigma -1)$ matrix over a binary field $GF(\varsigma =2)$. Consequently, a QC-LDPC code $C^{(1)}_{qc}$ of length $\mu_{r}(\varsigma-1)$, defined over the null space of $\mathbf{H}_{b}^{(1)}(\lambda_{c}, \mu_{r})$, is obtained with minimum distance and rate equal to $2f+1$ and $(\mu_{r}-\lambda_{c})/\mu_{r}$, respectively. Furthermore, QC-LDPC codes for various rate and lengths can be constructed for different choices of $\lambda_{c}$ and $\mu_{r}$. Thus, the proposed $(\Omega, f, \chi;\Xi)$-CBSEC construction gives QC-LDPC codes with flexible code length and rate. 
 \subsection{CBSEC-Based Joint Design of QC-LDPC Codes}
 Here, a joint framework of QC-LDPC codes is provided using the CBSEC-based construction of QC-LDPC codes constructed in Section IV-B. 
 Suppose that the null space of three parity-check matrices $\mathbf{H}_{J_{1}\times N}^{(1)}$, $\mathbf{H}_{J_{2}\times N}^{(2)}$, and $\mathbf{H}_{J_{3}\times N}^{(3)}$, realizing the source and relay nodes, give three QC-LDPC codes $C^{(1)}(N,J_{1})$, $C^{(2)}(N,J_{2})$, and $C^{(3)}(N,J_{3})$, respectively. Moreover, $\mathbf{H}_{J_{1}\times N}^{(1)}$, $\mathbf{H}_{J_{2}\times N}^{(2)}$, and $\mathbf{H}_{J_{3}\times N}^{(3)}$
 are obtained from $(\Omega, f, \chi;\Xi)$-CBSEC construction of QC-LDPC codes provided in (\ref{46}). Thus, the joint construction of parity-check matrix $\mathbf{H}_{(J_{1}+J_{2}+J_{3})\times (N+J_{3})}$ for the joint decoding of $U_{f}$ signal can be expressed as follows  
   \begin{align}
 \mathbf{H}_{(J_{1}+J_{2}+J_{3})\times (N+J_{3})}=
 	\begin{bmatrix}
 		\mathbf{H}_{J_{1}\times N}^{(1)}& \mathbf{0}_{J_{1}\times J_{3}} \\ 
 		\mathbf{H}_{J_{2}\times N}^{(2)} & \mathbf{0}_{J_{2}\times J_{3}}\\ 
 		\mathbf{H}_{J_{3}\times N}^{(3)}& \mathbf{H}_{J_{3}\times J_{3}}^{(3)}
 	\end{bmatrix},
 	\label{47}
 \end{align}
where $J_{1}=\lambda_{c1}(\varsigma -1)$, $J_{2}=\lambda_{c2}(\varsigma -1)$,  $J_{3}=\lambda_{c3}(\varsigma -1)$, and  $N=\mu_{r}(\varsigma -1)$, for $1\leq \lambda_{c1}, \lambda_{c2}, \lambda_{c3}\leq 2f$ and $1\leq \mu_{r}\leq wf$. Finally, at the destination node, the joint parity-check matrix $\mathbf{H}_{(J_{1}+J_{2}+J_{3})\times (N+J_{3})}$ is used to realize the SPA-based joint iterative decoder which decodes the corrupted streams of $U_{f}$ data revived in two distinct time frames for considered BSC-enabled cooperative NOMA system.  
%

\section{Numerical Results and Discussion}
To analyze the performance of the proposed BSC-enabled cooperative NOMA system, the simulation results have been presented in terms of energy efficiency under various performance parameters. For comparison, the performance of the proposed cooperative NOMA framework with backscatter tag (WBST) is compared with the cooperative NOMA network with no backscatter tag (NBST)\footnote{It is difficult to find a suitable benchmark for comparison purpose due to the novelty of proposed framework. Therefore, the performance of the proposed BSC cooperative NOMA network is compared with its simplified cooperative NOMA network with no backscatter tag.}. Unless specified, the simulation parameters adopted for considered systems are summarized in TABLE \ref{TABLE 1}.

\begin{figure*}
	\centering
	\begin{minipage}[b]{.4\textwidth}
		\includegraphics [width=0.95\textwidth]{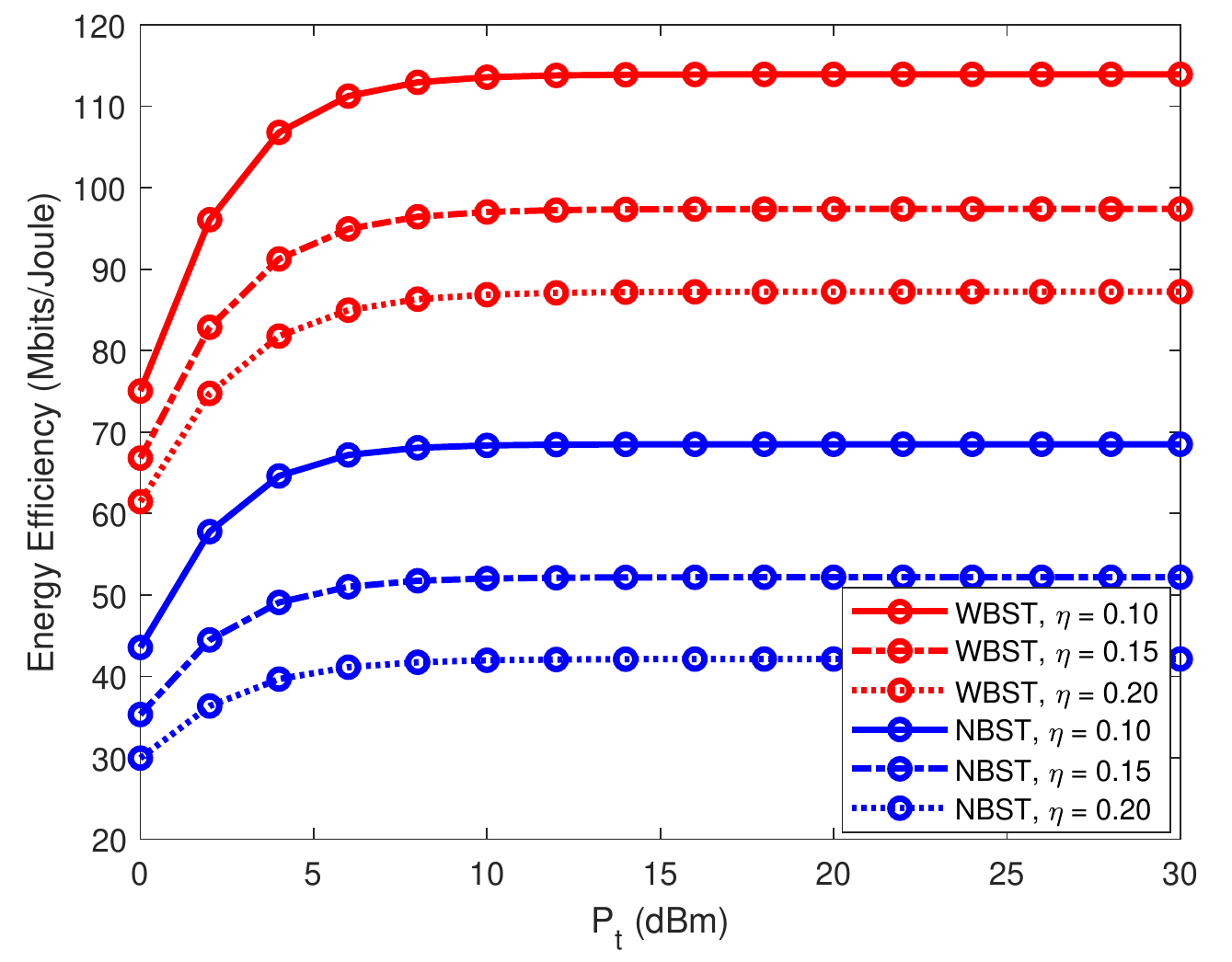}
		\caption{The effect of maximum available transmit power of source on the energy efficiency under different values of $\eta$.}
		\label{f2}
	\end{minipage}\qquad
	\begin{minipage}[b]{.4\textwidth}
		\includegraphics [width=0.95\textwidth]{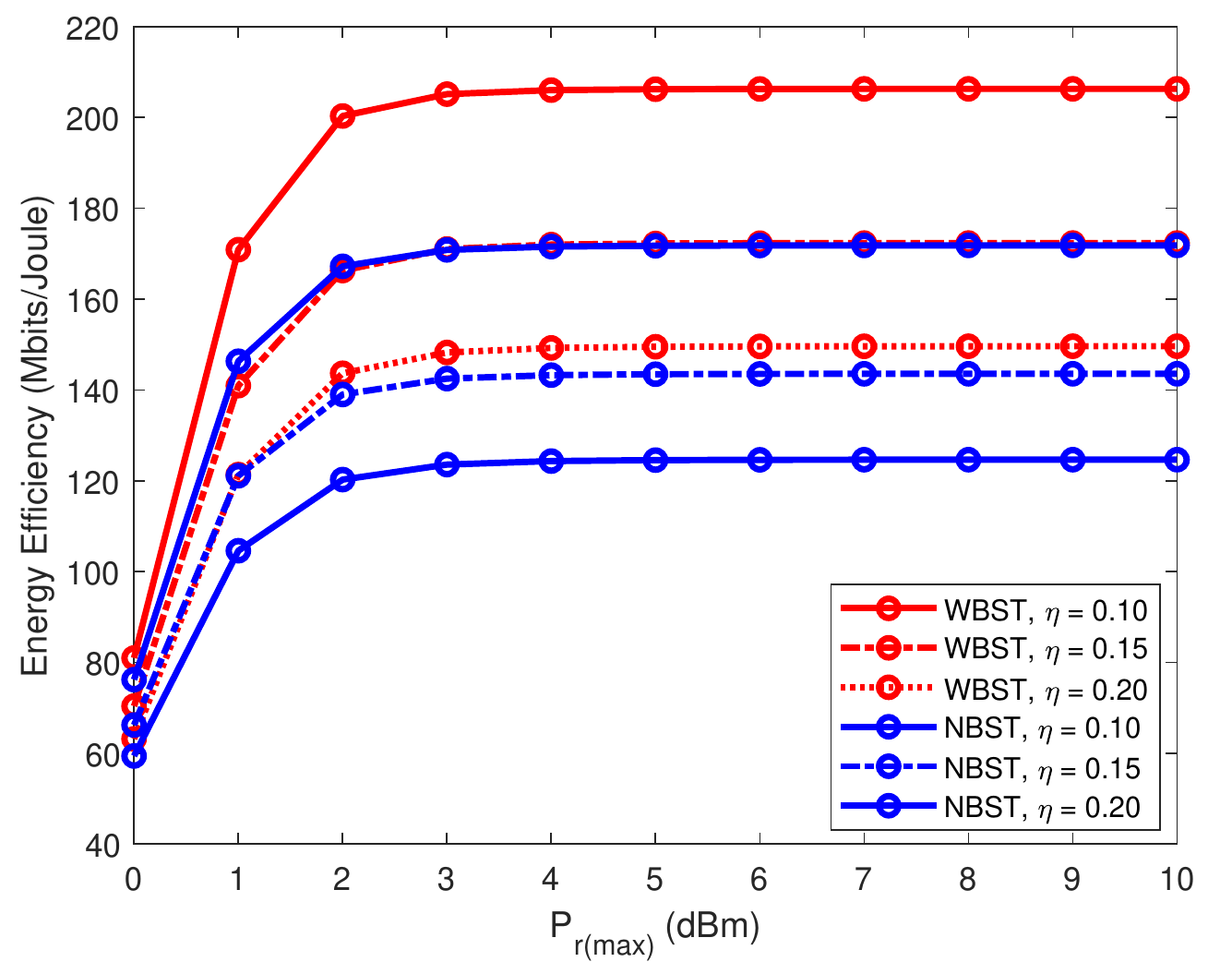}
		\caption{The impact of maximum available power of relay node on the energy efficiency of the system under different values of $\eta$.}
		\label{f3}
	\end{minipage}
\end{figure*}

The effect of total available transmit power of source node $P_{t}$ on the energy efficiency of the considered BSC-enabled cooperative NOMA system under different values of imperfect SIC parameter $\eta$ has been depicted in Fig. \ref{f2}. It can be observed that initially, the energy efficiency of the system increases by increasing the value $P_{t}$. However, after a certain point, a further increment in the value of $P_{t}$ has no impact on the performance of the system, and the energy efficiency becomes constant. The reason for this trend is that the transmit power becomes efficient to meet QoS requirement and allocated power remains unchanged with further increase in the value of $P_{t}$. Also, Fig. \ref{f2} shows that the energy efficiency of the system decreases for the higher values of $\eta$. Furthermore, the simulation results demonstrate that the proposed WBST system outperforms its NBST competitor in terms of energy efficiency under different values of $\eta$.

 Fig. \ref{f3} depicts the impact of total available transmit power of relay node $P_{r(max)}$ on the energy efficiency of the system for different values of $\eta$. Initially, the energy efficiency of the system increases by increasing the total available power of the relay node $P_{r(max)}$. However, energy efficiency becomes constant after a certain point, because the relay power is efficient to meet the minimum rate requirement of NOMA users and allocated relay power remains unchanged with a further increase in the value of $P_{r(max)}$. It can also be seen that energy efficiency decreases for the higher values of $\eta$. The performance of the proposed WBST system has been compared with its NBST counterpart under different values of $\eta$. Further, simulation results evince that the proposed WBST outperforms its NBST counterpart in terms of the energy efficiency of the system.       

\begin{figure*}
	\centering
	\begin{minipage}[b]{.4\textwidth}
		\includegraphics [width=0.95\textwidth]{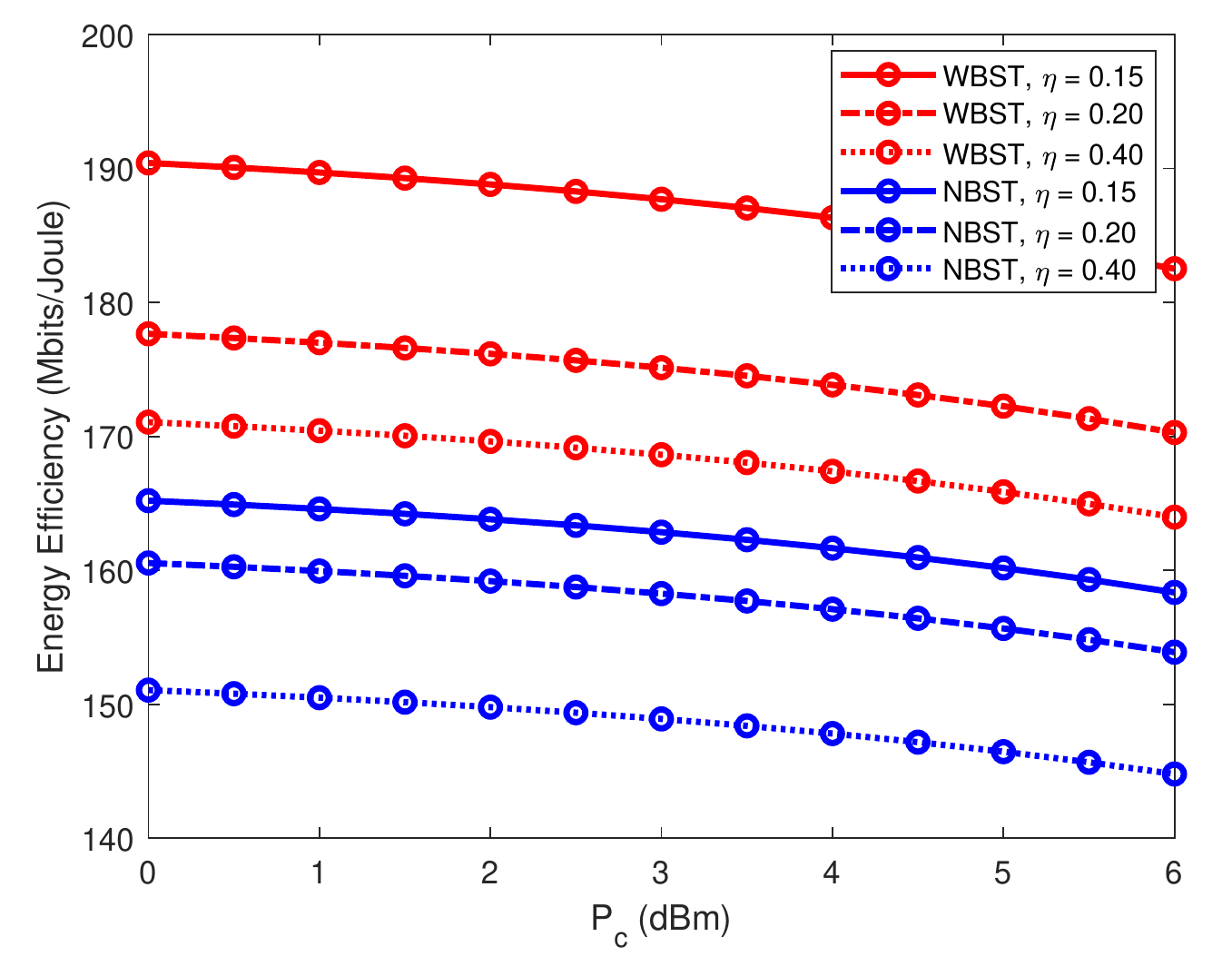}
		\caption{The impact of increasing $P_{c}$ on energy efficiency under different values of $\eta$.}
		\label{f4}
	\end{minipage}\qquad
	\begin{minipage}[b]{.4\textwidth}
		\includegraphics [width=0.95\textwidth]{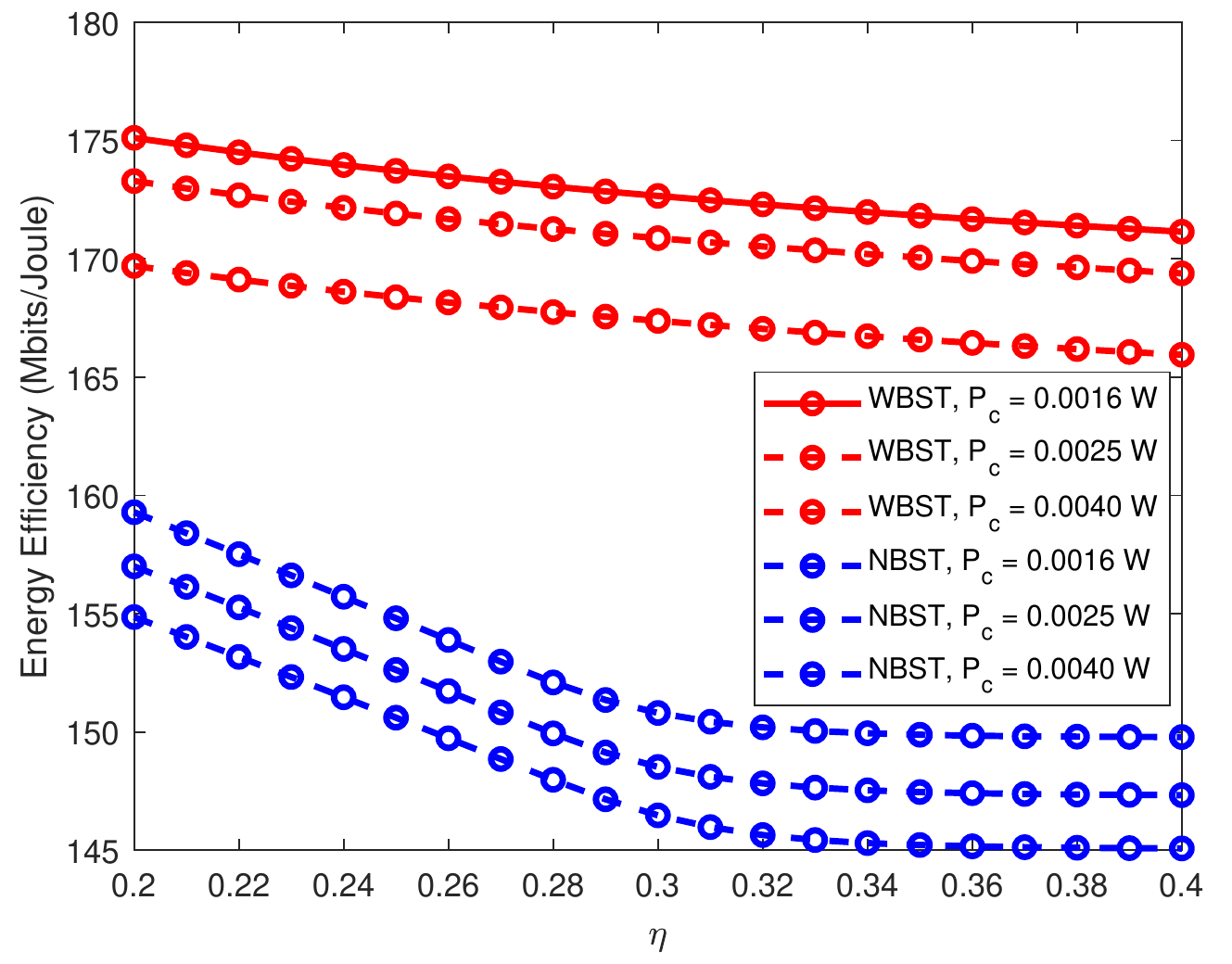}
		\caption{The impact of increasing $\eta$ on energy efficiency under different values of $P_{c}$.}
		\label{f5}
	\end{minipage}
\end{figure*}

The effect of circuit power $P_{c}$ on the energy efficiency of the system under different values of imperfect SIC parameter $\eta$ is demonstrated Fig. \ref{f4}. It can be seen that the energy efficiency of the system decreases by increasing the values of $P_{c}$. Since, an increment in the value of $P_{c}$ increases the total power consumption of the system, which ultimately decreases the energy efficiency of the system. Further, the simulation results have been presented for different values of $\eta$, i.e., 0.15, 0.20, and 0.40. We can observe that higher energy efficiency is achieved for lower values of $\eta$. Furthermore, the analysis of simulation results reveals that the proposed WBST system outperforms its NBST counterpart in terms of energy efficiency under different values of $\eta$.
\begin{figure*}
	\centering
	\begin{minipage}[b]{.40\textwidth}
		\includegraphics [width=0.97\textwidth]{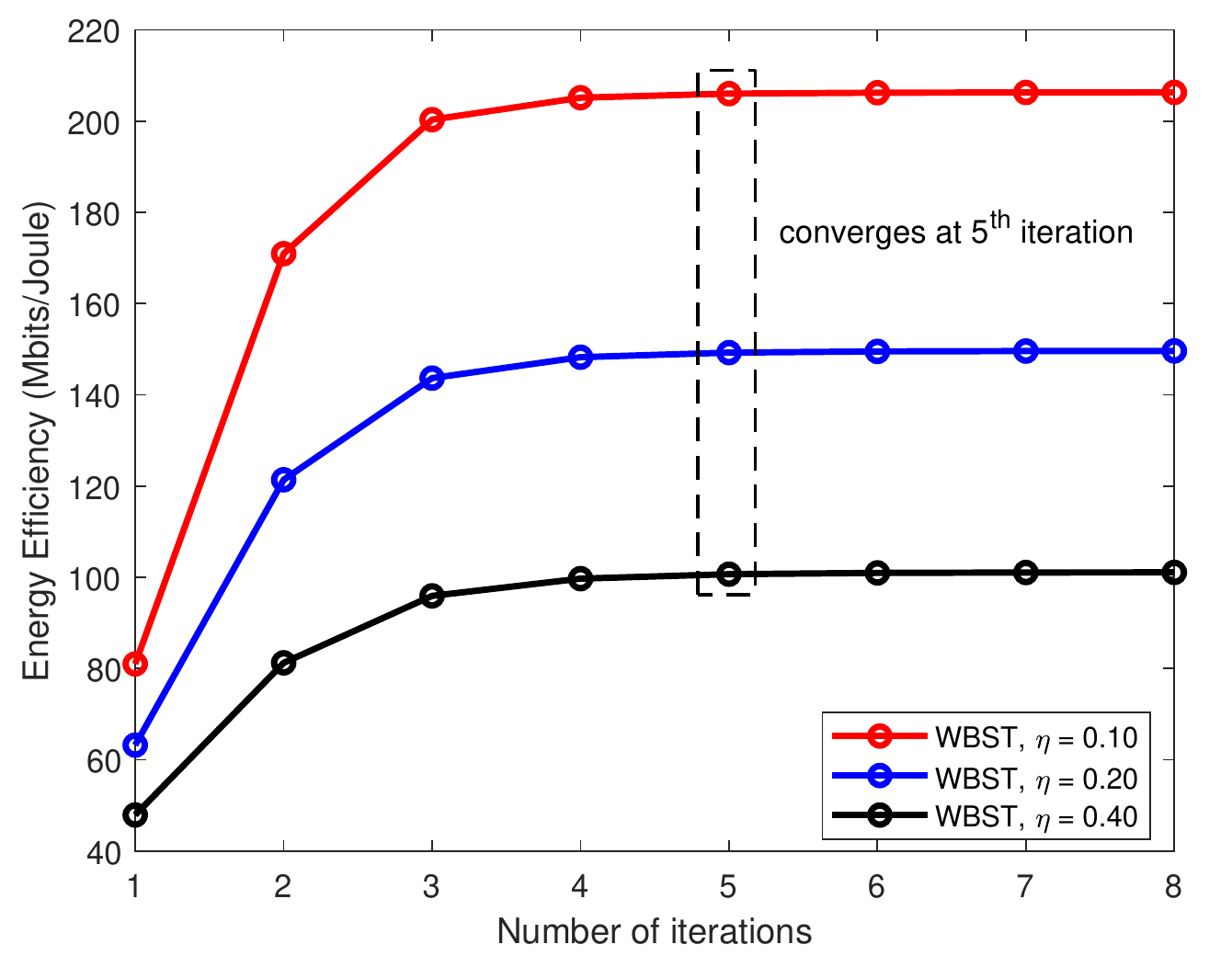}
		\caption{The convergence of energy efficiency for different number of iterations}
		\label{f6}
	\end{minipage}\qquad
	\begin{minipage}[b]{.40\textwidth}
		\includegraphics [width=0.95\textwidth]{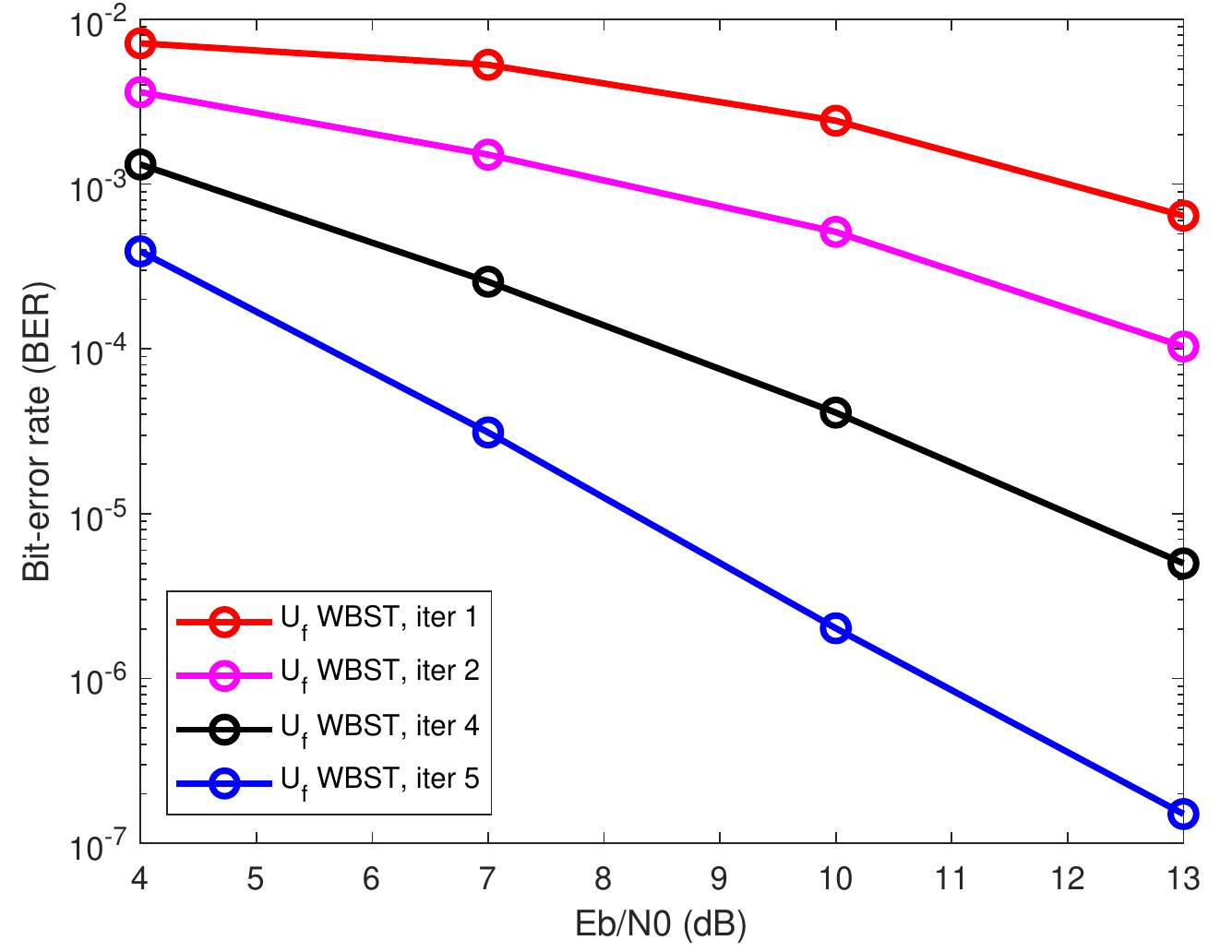}
		\caption{BER performance analysis of proposed jointly-designed QC-LDPC codes under different decoding iterations.}
		\label{f7}
	\end{minipage}
\end{figure*}

The impact of imperfect SIC parameter $\eta$ on the energy efficiency of considered BSC-enabled cooperative NOMA system has been depicted in Fig. \ref{f5}. It can be seen that the energy efficiency of the system decreases with the increasing values of $\eta$. This performance degradation of the system occurs due to the interference caused by the far user while decoding the near user data during the first time slot of the BSC cooperative NOMA system. Specifically, more interference is faced by $U_{n}$ for higher values of $\eta$. Consequently, more resources are required to meet the QoS requirement of $U_{n}$ which ultimately decreases the energy efficiency of the system. Further, Fig. \ref{f5} also shows that the energy efficiency decreases by increasing the circuit power $P_{c}$. Moreover, simulation results evince that the proposed WBST system outperforms its NBST competitor by providing excellent performance in terms of the energy efficiency of the system.
   
   
\begin{table*}
	\captionsetup{justification=centering}
	\caption{Simulation Parameters}
	\label{TABLE 1}
	\centering
	\begin{tabular}{| c | c |}
		\hline
		{\textbf{Parameters}} & {\textbf{Value}} \\
		\hline
    	Imperfect SIC parameter ($\eta$) & 0.1 $-$ 0.4 \\
		\hline
		Bandwidth $(W)$ & 1 MHz \\
		\hline
		Path-loss exponent & 4 \\
		\hline
		Transmit power budget &  30 dBm \\
		\hline
		Minimum data rate for QoS ($R_{min}$) & 0.5 Mbits/sec \\\hline
		Relay power budget & 10 dBm \\
		\hline
		Circuit power ($P_{c}$) & 0.001 W \\
		\hline
		Noise power ($\sigma^{2}$) &  -$114$ dBm\\
		\hline
		Fast-fading & Rayleigh-fading \\
		\hline
	\end{tabular}
\end{table*}

The convergence behavior of the considered BSC cooperative NOMA system in terms of energy efficiency under different values of $\eta$ has been depicted in Fig. \ref{f6}. It can be observed that the number of iterations required for the convergence are same for all values of $\eta$. For instance, only 5 iterations are required for convergence when $\eta$=0.1, 0.2, and 0.4. Consequently, it can be concluded that $\eta$ affects the system by decreasing its performance in terms of energy efficiency, however, its effect on the convergence of the system is negligible. 

Furthermore, the far user of considered BSC cooperative NOMA system suffers from performance degradation due to indirect communication link with BS and bad channel conditions. Therefore, to improve the performance of $U_{f}$, in terms of BER, the far user's data is jointly decoded by a SPA-based joint iterative decoder realized by jointly-designed QC-LDPC codes given by Eq. (\ref{47}). The simulation results in terms of the error-correction performance of $U_{f}$ under different decoding iterations of SPA-based joint iterative decoder have been presented in Fig. \ref{f7}. Simulation results evince that the proposed jointly-designed QC-LDPC codes provide an excellent BER performance to jointly decode the $U_{f}$ data for considered BSC-enabled coded cooperative NOMA system with only a few decoding iterations under Rayleigh-fading transmission.

\section{Conclusion and Remarks}
A novel alternating optimization framework has been proposed to enhance the energy efficiency of considered BSC-enabled cooperative NOMA system for upcoming next-generation communication networks. More specifically, we aim to maximize the energy efficiency of the system by optimizing the transmit power of the source, power allocation coefficients, and power of the relay node under the imperfect SIC decoding at the receiver. Further, the energy efficiency of considered backscatter-enabled cooperative NOMA system is maximized by considering QoS, power budgets, and cooperation constraints to meet the minimum rate requirements for NOMA users. The proposed optimization problem was decoupled into three sub-problems to find the optimal solutions which finally yields an energy-efficient low complexity algorithm with only a few iterations for convergence. Furthermore, a class of $(\Omega, f, \chi;\Xi)$-CBSEC based jointly designed QC-LDPC codes has been proposed to enhance the performance of far user for considered BSC-enabled coded-cooperative NOMA system. Moreover, the simulation results evince that the proposed WBST system outperforms its NBST counterpart by providing an efficient performance in terms of energy efficiency. Also, proposed jointly-designed QC-LDPC codes provide an efficient error-correction performance, in terms BER, by jointly decoding the far user data for considered BSC cooperative NOMA system with only a few decoding iterations. Consequently, efficient channel coding techniques could play a crucial role to tackle the chain process of error propagation due to imperfect SIC decoding for NOMA-enabled next-generation communication systems.      

\bibliographystyle{IEEEtran}
\bibliography{Wali_Ref}

\begin{IEEEbiography}
	[{\includegraphics[width=1in,height=1.5in,clip,keepaspectratio]{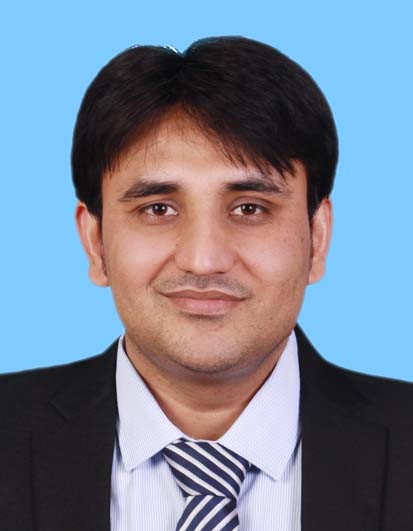}}]{MUHAMMAD ASIF} 
	was born in Rahim Yar Khan, Bahawalpur Division, Pakistan, in 1990. He received the Bachelor of Science (B.Sc) degree in Telecommunication Engineering from The Islamia University of Bahawalpur (IUB), Punjab, Pakistan, in 2013, and Master degree in Communication and Information Systems from Northwestern Polytechnical University (NWPU), Xian, Shaanxi, China, in 2015. He also received Ph.D. degree in Information and Communication Engineering from University of Science and Technology of China (USTC), Hefei, Anhui, China in 2019. Currently, Dr. Asif is working as a post-doctoral researcher at the Department of Electronics and Information Engineering in Shenzhen University, Shenzhen, Guangdong, China. He has authored/co-authored several journal and conference papers. His research interests include Wireless Communication, Channel Coding, Coded-Cooperative Communication, Optimization and Resource Allocation, Backscatter-Enabled Wireless Communication, IRS-Assisted Next-generation IOT Networks.
\end{IEEEbiography}

\begin{IEEEbiography}
	[{\includegraphics[width=1in,height=1.25in,clip,keepaspectratio]{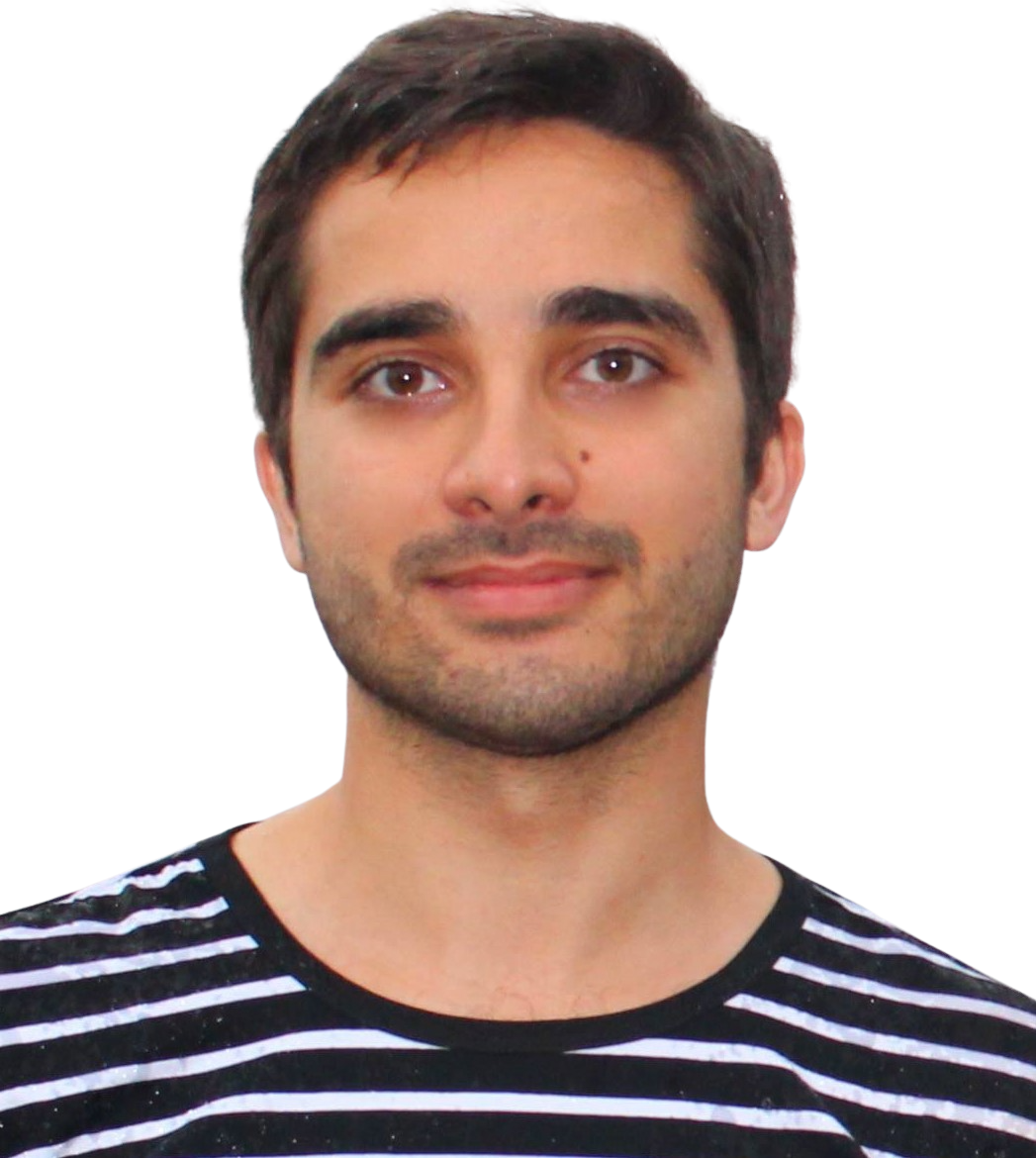}}]{Asim Ihsan}
	received the master’s degree in information and communication engineering from Xian Jiaotong University, Xi’an, China, and the Ph.D.degree in information and communication engineer-ing from Shanghai JiaoTong University, Shanghai,China. He is currently working as a Post-Doctoral Research Officer with the School of Computer Science and Electronic Engineering, Bangor University,U.K. He is also a Global Talent Visa Holder of U.K. His research interests include energy-efficient resource allocations for beyond 5G wireless communication technologies through convex/non-convex optimizations and machine learning.
\end{IEEEbiography}

\begin{IEEEbiography}
	[{\includegraphics[width=1in,height=1.5in,clip,keepaspectratio]{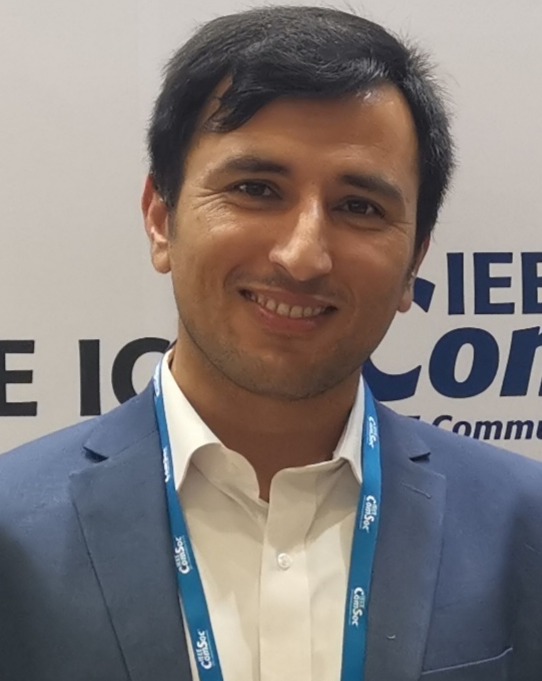}}]{Wali Ullah Khan} (Member, IEEE)
	received the Master degree in Electrical Engineering from COMSATS University Islamabad, Pakistan, in 2017, and the Ph.D. degree in Information and Communication Engineering from Shandong University, Qingdao, China, in 2020. He is currently working with the Interdisciplinary Centre for Security, Reliability and Trust (SnT), University of Luxembourg, Luxembourg. He has authored/coauthored more than 50 publications, including international journals, peer-reviewed conferences, and book chapters. His research interests include convex/non-convex optimizations, non-orthogonal multiple access, reflecting intelligent surfaces, ambient backscatter communications, Internet of things, intelligent transportation systems, satellite communications, physical layer security, and applications of machine learning.
\end{IEEEbiography}

\begin{IEEEbiography}
	[{\includegraphics[width=1in,height=1.25in,clip,keepaspectratio]{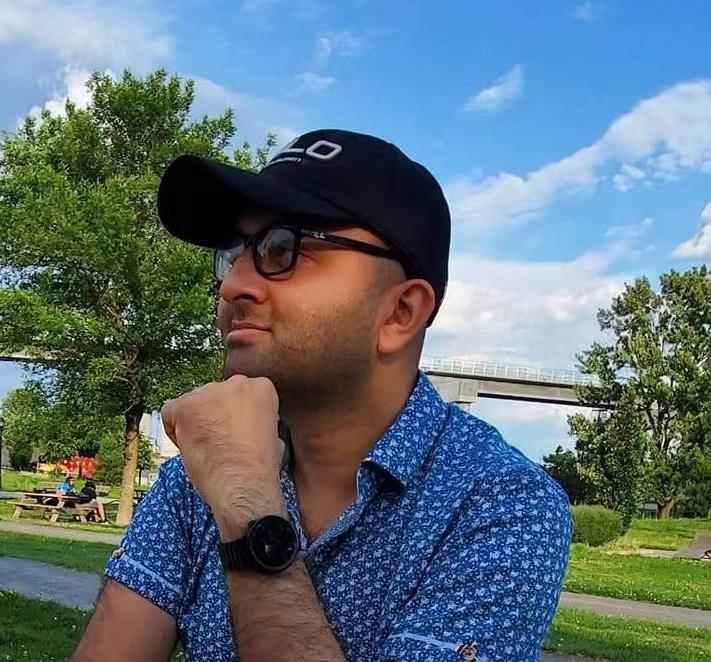}}]{Ali Ranjha}
 received his Ph.D. degree at Ecole de Technologie Superieure (ETS), Universite du Quebec, Montreal, Canada in January 2022, where he is also currently pursuing his postdoctoral research. In 2018, he completed his M.S. degree in innovation in telecommunications from Lancaster University, U.K. under a prestigious Higher Education Funding Council of England (HEFCE) bursary, His research interests span diverse areas, such as fundamental communication theory, unmanned aerial vehicle (UAV) communications, internet of things (IoT), ultra-reliable and low latency communications (URLLC), and optimization in resource-constrained networks.
\end{IEEEbiography}

\begin{IEEEbiography}
	[{\includegraphics[width=1in,height=1.25in,clip,keepaspectratio]{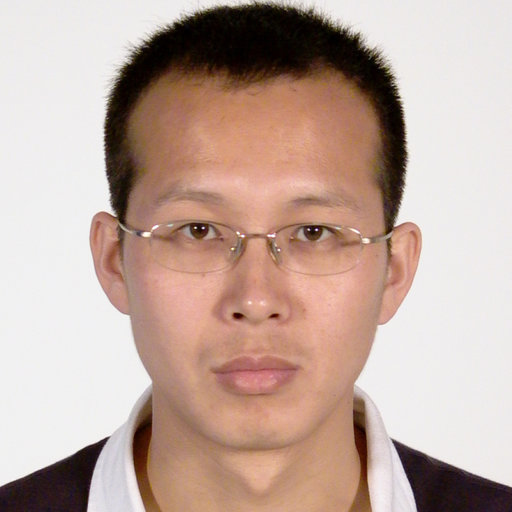}}]
	{Shengli Zhang} (Senior Member, IEEE) received
	the B.Eng. degree in electronic engineering and the
	M.Eng. degree in communication and information
	engineering from the University of Science and
	Technology of China, Hefei, China, in 2002 and
	2005, respectively, and the Ph.D. degree from the
	Department of Information Engineering, Chinese
	University of Hong Kong, Hong Kong, in 2008.
	He joined the Communication Engineering
	Department, Shenzhen University, Shenzhen, China,
	where he is currently a Full Professor. From
	March 2014 to March 2015, he was a Visiting Associate Professor with
	Stanford University, Stanford, CA, USA. He is the pioneer of Physical-Layer
	Network Coding. He has published over 20 IEEE top journal papers and
	ACM top conference papers, including IEEE JOURNAL ON SELECTED
	AREAS IN COMMUNICATIONS, IEEE TRANSACTIONS ON WIRELESS
	COMMUNICATIONS, IEEE TRANSACTIONS ON MOBILE COMPUTING,
	IEEE TRANSACTIONS ON COMMUNICATIONS, and ACM Mobicom. His
	research interests include blockchain, physical layer network coding, and
	wireless networks.
	Prof. Zhang severed as an Editor for IEEE TRANSACTIONS ON
	VEHICULAR TECHNOLOGY, IEEE WIRELESS COMMUNICATIONS
	LETTERS, and IET Communications. He has also severed as a TPC member
	in several IEEE conferences.
\end{IEEEbiography}

\begin{IEEEbiography}
	[{\includegraphics[width=1in,height=1.25in,clip,keepaspectratio]{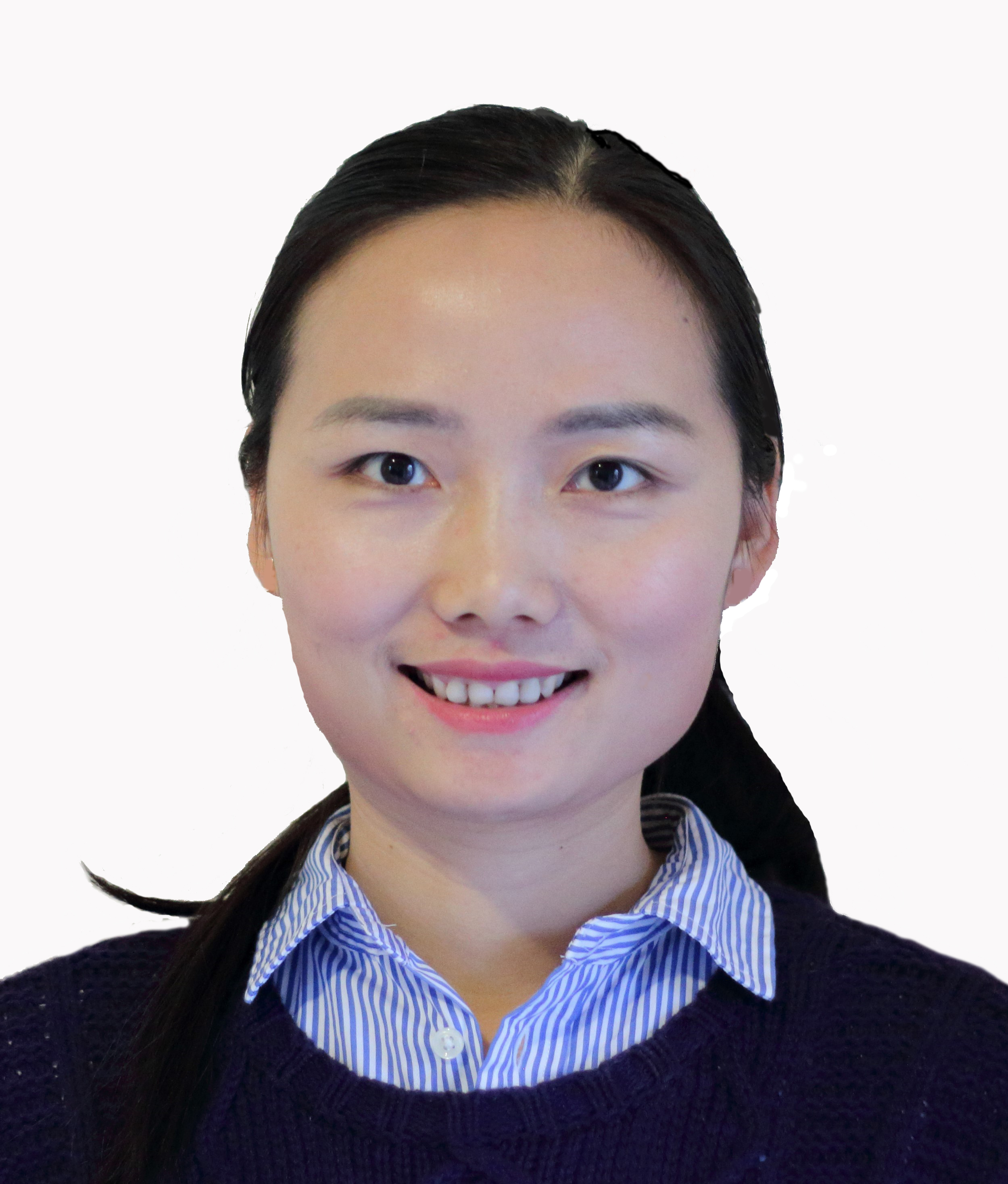}}]
	{Sissi Xiaoxiao Wu} (Member, IEEE) received the B.Eng. degree in electronic information engineering from the Huazhong University of Science and Technology, Wuhan, China, in 2005, the M.Phil. degree from the Department of Electronic and Computer Engineering, Hong Kong University of Science and Technology, Hong Kong, in 2009, and the Ph.D. degree in electronic engineering from the Chinese University of Hong Kong (CUHK), Hong Kong, in 2013.,From December 2013 to November 2015, she was a Postdoctoral Fellow in the Department of Systems Engineering and Engineering Management, CUHK. From December 2015 to March 2017, she was a Postdoctoral Fellow in the Signal, Information, Networks and Energy Laboratory supervised by Prof. A. Scaglione of Arizona State University, Tempe, AZ, USA. She is now an Associate Professor at the Department of Communication and Information Engineering, Shenzhen University, Shenzhen, China. Her research interests are in wireless communication theory, optimization theory, stochastic process, and channel coding theory, and with a recent emphasis on the modeling and data mining of opinion diffusion in social networks. She is now an Associate Editor of IEEE Transactions on Vehicular Technology and serving as an IEEE Signal Processing Society SAM Technical Committee Member.
\end{IEEEbiography}

\end{document}